
\magnification=1200
\hsize=15truecm
\vsize=23truecm
\baselineskip 20 truept
\voffset=-0.5truecm
\parindent=0cm
\overfullrule=0pt

\def\pmb#1{\leavevmode\setbox0=\hbox{$#1$}\kern-.025em\copy0\kern-\wd0
\kern-.05em\copy0\kern-\wd0\kern-.025em\raise.0433em\box0}

\def\Ai{\hbox{\hbox{${\cal A}$}}\kern-1.9mm{\hbox{${/}$}}}
\def\Vi{\hbox{\hbox{${\cal V}$}}\kern-1.9mm{\hbox{${/}$}}}
\def\Di{\hbox{\hbox{${\cal D}$}}\kern-1.9mm{\hbox{${/}$}}}
\def\lam{\hbox{\hbox{${\lambda}$}}\kern-1.6mm{\hbox{${/}$}}}
\def\D{\hbox{\hbox{${D}$}}\kern-1.9mm{\hbox{${/}$}}}
\def\A{\hbox{\hbox{${A}$}}\kern-1.8mm{\hbox{${/}$}}}
\def\V{\hbox{\hbox{${V}$}}\kern-1.9mm{\hbox{${/}$}}}
\def\parz{\hbox{\hbox{${\partial}$}}\kern-1.7mm{\hbox{${/}$}}}
\def\B{\hbox{\hbox{${B}$}}\kern-1.7mm{\hbox{${/}$}}}
\def\R{\hbox{\hbox{${R}$}}\kern-1.7mm{\hbox{${/}$}}}
\def\si{\hbox{\hbox{${\xi}$}}\kern-1.7mm{\hbox{${/}$}}}

\centerline{\bf BOSONIZATION OF FERMI SYSTEMS IN}

\centerline{\bf ARBITRARY DIMENSION IN TERMS OF GAUGE FORMS}

\vskip 0.5truecm

\centerline{J. Fr\"ohlich$^1$, R. G\"otschmann$^1$, P.A. Marchetti$^{2*}$}

\vskip 0.3truecm
1 -- Theoretical Physics, ETH--H\"onggerberg, CH--8093 Z\"urich
\vskip 0.2truecm
2 -- Dipartimento di Fisica,  Universit\`a di Padova and
I.N.F.N.

\qquad
Sezione di Padova, I - 35131,  Italy
\vskip 1truecm
{\bf Abstract.} We present a general method to bosonize systems of Fermions
with infinitely many degrees of freedom, in particular systems of
non-relativistic electrons at positive density, by expressing the
quantized conserved electric charge- and current density in terms of a
bosonic antisymmetric tensorfield of a rank d--1, where d is the
dimension of space. This enables us to make concepts and tools from gauge
theory available for the purpose of analyzing electronic structure
of non-relativistic matter. We apply our bosonization identities and
concepts from gauge theory, such as Wegner -'t Hooft duality, to a
variety of systems of condensed matter physics: Landau-Fermi liquids, Hall
fluids, London superconductors, etc.. Among our results are an exact
formula for the plasmon gap in a metal, a simple derivation of the
Anderson-Higgs mechanism in superconductors, and an analysis of the
orthogonality catastrophe for static sources.

\vskip 1.5truecm
\centerline{DFPD 94/TH/36}

\centerline{May 1994}
\vskip1.5truecm
(*) Supported in part by M.P.I. This work is carried out in the framework
of the European Community Research Programme ``Gauge Theories, applied
supersymmetry and quantum gravity" with a financial contribution under
contract SC1--CT92--D789.
\vfill\eject

{\bf 1. \ Introduction and summary of main results}
\vskip 0.5truecm
In this paper, we develop a conceptual framework, based on bosonization
of quantum systems with infinitely many degrees of freedom, which we
expect to be useful in attempts to classify states of non-relativistic
matter at very low temperatures. In this paper, we focus our attention
on the analysis of electronic structure. Magnetic properties will be
discussed in a separate paper.

The basic ideas underlying our approach are very simple: our starting point
is to study the response of a quantum system of charged particles to
perturbations by external electro-magnetic fields. Thus we couple the
electric current density to an arbitrary, smooth external
electromagnetic vector potential, $A$, and then attempt to
calculate the partition function, $\Xi(A)$, of the system as a
functional of $A$.

Of course, this is a very complicated task. However, in order to
classify electronic structure of non-relativistic matter, we
are really only interested in understanding the behaviour of
the \underbar{effective action},

$$
S (A) \equiv  -i \bar h {\rm ln} \Xi (A) \eqno(1.1)
$$

on very large distance scales and at very low
frequencies. We thus study families of systems confined
to ever larger cubes, $\Omega_\lambda : = \{{\bf x} :
{{\bf x} \over \lambda} \in \Omega \}$, in physical space
${\bf R}^d, d = 1, 2, 3$ where $\Omega$ is a fixed cube in ${\bf R}^d$
and $1 \leq \lambda < \infty$ is a scale parameter. We keep the
particle density, $\rho$,  and the temperature $T (\approx 0)$
constant. We then couple the electric current density of
the system confined to $\Omega_\lambda$ to a vector
potential $A^{(\lambda)}$ given by

$$
A^{(\lambda)} (t, {\bf x}) \equiv \lambda^{-1} A \Bigl({t \over
\lambda},{{\bf x}  \over \lambda}\Bigr), {{\bf x} \over \lambda} \in \Omega,
$$

where $A$ is an arbitrary , but $\lambda$-independent vector potential on
space-time ${\bf R} \times \Omega$. We then study the behaviour of
$S^{\Omega_\lambda} (A^{(\lambda)})$ when $\lambda$ becomes large.

More precisely, we attempt to expand $S^{\Omega_\lambda}
(A^{(\lambda)})$ in powers of $\lambda^{-1}$ around $\lambda = \infty$
(to a finite order) and define the \underbar{scaling limit}, $S^\star
(A)$, of the effective action to be the \underbar{coefficient of the}
\underbar{leading power} of $\lambda$ in that expansion, (in the limit
when $\Omega \nearrow {\bf R}^d$); see [1].

One of our main contentions in this paper is the ``quasi--theorem" that
$S^\star (A)$ is \underbar{quadratic} in $A$, for all systems of
non-relativistic electrons at positive density we know of; in
particular for insulators, Landau Fermi liquids, metals, (incompressible)
quantum Hall fluids, superconductors.

We call this claim a ``quasi-theorem", because we do not know a
proof of it that would cover all imaginable cases. We thus rely on a
case-by-case analysis. The detail of our analysis will appear  in
separate papers, [2]. It should be emphasized, at this point, that, except
in the case of insulators and incompressible quantum Hall fluids, the proof
of the ``quasi-theorem" is never trivial and does \underbar{not}
simply follow from dimensional analysis.

It is well know that $S (A)$ is the generating  functional
of connected Green functions of the electric current density and that it is
\underbar{gauge-invariant}, i.e.,

$$
S(A + d \chi) = S (A), \eqno(1.2)
$$

where $\chi$ is an arbitrary real function and $d \chi$ is its
gradient\footnote*{We shall use notations and elementary notions of
Cartan's exterior calculus, throughout this paper}
Clearly, gauge-invariance persists up on passing to the scaling
limit. Using our ``quasi-theorem", we conclude that

$$
S^\star (A) = {1 \over 2}  (A, \Pi^\star A)
= {1 \over 2} \int d^{d+1} x \int d^{d+1} y A_\mu (x)
\Pi^{\star\mu\nu} (x - y) A_\nu (y), \eqno(1.3)
$$

where, for $x \not = y$, $\Pi^{\star\mu\nu} (x - y)$ is given by the scaling
limit of the two-current Green function $<T \Bigl({\cal J}^\mu (x)
{\cal J}^\nu (y)\Bigr) >^c$.

By gauge invariance, or, equivalently, current conservation,

$$
\partial_\mu \Pi^{\star\mu\nu} = \partial_\nu \Pi^{\star\mu\nu} = 0.
\eqno(1.4)
$$

Furthermore, $\Pi^{\star\mu\nu}$ inherits all the symmetries of the system;
(in
(1.3), we have assumed translation invariance). Thus, classifying
electronic structure  of a system reduces, in the scaling limit and
under the assumption that the quasi-theorem holds, to a classification
of ``vacuum polarisation tensors", $\Pi^{\star\mu\nu}$, satisfying (1.4) and
having certain symmetries. This is a straightforward task; see Sect. 6
and [2].

In principle, essentially all information concerning electronic
properties of a system can be retrieved from its effective action $S
(A)$ by fairly straightforward calculations. However, in order to
evoke and then apply analogies with other physical systems, in particular
with gauge theories of elementary particle physics, it is useful to
embark on a detour. The detour chosen in this paper is
\underbar{bosonization}. The idea (see [3]) is as follows: since
the electric current density ${\cal J} = ({\cal J}^\mu)^d_{\mu=0}$ is
conserved, it can be derived from a ``potential",

$$
{\cal J}^\mu = \epsilon^{\mu\mu_1...\mu_d} \partial_{\mu_1}
b_{\mu_2...\mu_d}, \eqno(1.5)
$$

where $\epsilon$ is the totally antisymmetric $\epsilon$-tensor and
$b_{\mu_2,...,\mu_d}$ is an antisymmetric tensor field of rank $d-1$.
In the language of differential forms, eq.(1.5) is expressed as

$$
{\cal J} = ^* d b, \eqno(1.6)
$$

where $*$ is the Hodge $*$ operation and $d$ denotes exterior
differentiation. The ``potential" $b$ of the current density is
determined by (1.6) only up to the exterior derivation of an antisymmetric
tensor field of rank $d-2$, (a $(d-2)$-form). Thus $b$ is
what one calls a \underbar{``gauge form"}.

For one-dimensional systems, $b$ is a scalar and is determined by (1.6) up
to a constant. In two dimensions, $b$ is a 1-form determined by ${\cal J}$
up to the exterior derivative of an arbitrary 1-form.

The basic idea is then to deduce an effective field theory for the field
$b$ from $S (A)$. This field theory is determined by an action $\tilde
S (b)$. Choosing units in which $\hbar = 1$ and using an imaginary-time
(euclidean) formulation, $\tilde S(b)$ is obtained from $S (A)$ by
functional Fourier transformation

$$
e^{- \tilde S(b)} = N^{-1} \int e^{-S (A)} e^{{i \over 2 \pi}
\int A \wedge db} {\cal D} A, \eqno(1.7)
$$

where $N$ is a (divergent) normalization factor (proportional to the volume
of the Lie algebra of $U(1)$-gauge transformations). Gauge invariance
implies that

$$
\tilde S (b+ d \Lambda)= \tilde S(b), \eqno(1.8)
$$

for an arbitrary ($d-2$)-form $\Lambda$ $(d \geq 2)$

Our quasi-theorem then implies that the low-wave-vector, low-energy modes
of the \underbar{bosonic} field $b$ are \underbar{non-interacting}, i.e.,
the scaling limit of the system described in terms of the $b$-field has
a \underbar{quadratic} action, $\tilde S^\star(b)$, whose form is
constrained by its gauge invariance, eq. (1.8), and by the symmetries
of the system.

Essentially all quantities of interest in the original system
can be expressed in terms of quantities referring to the
$b$-field. For example, current Green functions (at imaginary time) are
given by expectations of products of the ``field strenght" $d b$
in the functional measure

$$
\Xi^{-1} e^{- \tilde S (b)} {\cal D} b.
$$

Green functions of electron creation-and annihilation operators
turn out to be proportional to expectations of \underbar{disorder}
\underbar{operators} of the dual theory formulated in terms of the
$b$-field.

If $\Sigma$ is an open subset of physical space, and ${\cal Q}_
\Sigma$ denotes the
operator measuring the total electric charge inside $\Sigma$ then

$$
{\rm exp} \ (i \ \alpha {\cal Q}_\Sigma) \propto {\rm exp} \ (i \ \alpha
\int_{\partial\Sigma} b), \ \alpha \in {\bf R}, \eqno(1.9)
$$

i.e., the charge operator can be reconstructed from operators analogous to
the \underbar{Wilson loops} of gauge theory. The operators in (1.9)
are \underbar{dual} to the disorder operators describing electron creation
and - annihilation, in the sense of Wegner - 't Hooft duality, [4,5].
This suggests that we can carry over the consequences of 't Hooft
duality from gauge theory to condensed matter physics. As a
consequence we mention that if, at zero temperature, the total
electric charge operator, ${\cal Q} = ``{\rm lim}_{\Sigma \nearrow {\bf R}^d}
{\cal Q}_\Sigma"$
is well defined on the Hilbert space of all physical states of
the system then disorder Green functions, e.g. the Green function
of an electron creation-and an annihilation operator, exhibit
strong spatial cluster decomposition properties, and conversely.
This applies to insulators, incompressible Hall fluids and
superconductors with (unscreened) Coulomb two-body repulsion.
Furthermore, if the field $b$ couples the ground state of the
system to a massless quasi-particle then the total charge
operator does not exist, because charge fluctuations are
divergent in the thermodynamic limit, as in metals and
mass less super superconductors.

For \underbar{two-dimensional} systems, both, $A$ and $b$,
are 1-forms defined up to gradients of scalar functions.
In this case, eq. (1.7) enables us to define a notion
of \underbar{duality}: a system 1 and a system 2 are
\underbar{dual} to each other iff

$$
\tilde S^\star_1 \propto \ S^\star_2. \eqno(1.10)
$$

It turns out that, in the sense of eq. (1.10), a two--dimensional
\underbar{insulator} is \underbar{dual} to a two--dimensional
\underbar{London superconductor}, a metal to a ``semi-conductor",
and an incompressible Laughin Hall fluid is
\underbar{self-dual}.

Besides the duality expressed in eqs. (1.7) and (1.10) there is
also a notion of Kramers-Wannier duality: in \underbar{any dimension}
d, a $U(1)$-gauge theory of an antisymmetric tensor field, $b$, of rank
$d-1$ is ``Kramers-Wannier-dual" to a \underbar{scalar} field
theory. Thus, for example, a London superconductor, corrected by
dynamical  Abrikosov vortices, is Kramers-Wannier dual to a
Landau-Ginsburg superconductor; see [1].

The two notions of duality sketched here are conceptually
quite clarifying and useful in a classification of electronic
properties of non-relativis
One of the principal advantages of reformulating the theory
of a system of electrons in terms of the tensor field $b$
(bosonization) is that this formulation is convenient to
explore systems obtained by perturbing a given one by
\underbar{two-body interactions}. A translation-invariant
two-body interaction, $I_{pert}$, has the form

$$
I_{pert} = \int d^{d+1} \times \int d^{d+1} y {\cal J}^\mu (x) V_{\mu\nu}
(x-y) {\cal J}^\nu (y) \eqno(1.11)
$$

After bosonization, the action of the perturbed system is
given by

$$
\tilde S_{tot} (b) = \tilde S(b) + \int d^{d+1}  x \int d^{d+1}
y (^* db)^\mu (x) V_{\mu\nu} (x-y) (^*db)^\nu (y), \eqno(1.12)
$$

where  $\tilde S (b)$ is the action of the  unperturbed
system. Note that, expressed in terms of the field $b$
the two-body interaction $I_{pert}$ is \underbar{quadratic}
(rather than quartic)! A conventional two-body
interaction described by an istantaneous two-body
potential corresponds to a kernel $V_{\mu\nu}$ given by

$$
V_{\mu\nu} (x - y) = \delta_{\mu 0} \delta_{\nu 0} V
({\bf x} - {\bf y}) \delta (x^0 - y^0), \eqno(1.13)
$$

with $x = (x^0, {\bf x}), y = (y^0, {\bf y})$.

Suppose now that the action of the perturbed system in the
scaling limit (scale parameter $\lambda \rightarrow \infty$),
$\tilde S^\star_{tot}$, is given by the scaling limit, $\tilde S^\star$,
of the action of the imperturbed system, perturbed by the long-range tail,
$I_{pert}^\star$, of the two-body interaction $I_{pert}$. From
our quasi-theorem, we infer that $\tilde S^\star$ is quadratic
in $b$, and hence, since $I^\star_{pert}$ is quadratic in $b$,
$\tilde S^\star_{tot}$ is \underbar{quadratic} in $b$, too, and, by
(1.8) and (1.12), gauge-invariant. It is given by

$$
\tilde S^\star_{tot} \sim \tilde S^\star + \lambda^\kappa I^\star_{pert},
\ (\lambda \rightarrow \infty),
\eqno(1.14)
$$

for some exponent $\kappa \geq 0$. It is plausible that  the assumption
that perturbation by $I$ and passage to the scaling limit are
commuing operations is justified if $V_{\mu\nu}$ is
\underbar{positive-definite} and of very \underbar{long range}
(e.g. Coulomb - Amp\`ere type).

In this case, our analysis yields a somewhat more precise form of the
``random phase approximation" (RPA). We apply these ideas to the
following systems:

(i) A metal perturbed by repulsive two-body
Coulomb interactions. In this case we obtain the exact
formula for the \underbar{plasmon gap}.

(ii) A massless London superconductor perturbed by repulsive two-body
Coulomb interactions. In this example we recover a precise
formulation of the \underbar{Anderson-Higgs}
\underbar{mechanism}.

(iii) A Landau-Fermi liquid perturbed by repulsive
two-body interactions, in the presence of a static source.
For a two-dimensional system of this type with Coulomb-Amp\`ere
interaction we discuss
a  possible cross over to a non-Fermi liquid behaviour
(\underbar{``Luttinger liquid"}).

We conclude this introduction with a brief summary of the contents of this
paper.

In Sect. 2, we recall some notions in the theory of differential forms
and fibre bundles and outline the theory of \underbar{gauge forms} of
which the potential $b$ of the conserved electric current density ${\cal J}$,
see eqs. (1.5) and (1.6), is a special case.

In Sect. 3, we derive our general \underbar{bosonization method},
based on eqs. (1.2), (1.6), (1.7) and (1.8), and present the
main identities it provides.
In Sect. 4, we introduce  \underbar{disorder fields} of the
bosonized theory and the \underbar{local electric charge operators}
${\cal Q}_{\Sigma}$. We then explain the connection between disorder
fields and electron creation- and annihilation operators. We
recall what is meant by 't Hooft duality and discuss its
implications.

As an application of the general theory developed in Sects. 2, 3
and 4, we discuss systems of relativistic massless fermions in one
space dimension, in Sect. 5. In this case, we recover the
standard identities of one - dimensional, abelian bosonization.
In condensed matter physics, these systems describe Landau -
and Luttinger Fermi liquids and can be studied using techniques from
chiral $U(1)$-current algebra.

In one dimension, it is not hard to extend our methods to an
analysis of magnetic properties. This is done by coupling the
spin degrees of freedom of the electrons to external non-abelian
gauge fields with gauge group given by $SU(2)$, see [1].
Applying our bosonization methods to this speciale case, we would
recover the formulas of non-abelian bosonization and of
$SU(2)$-current algebra, but we shall not present these results
(which actually have just appeared in a recent preprint by
Burgess and Quevedo [6]).

In Sect. 6, we apply our methods to systems from condensed matter
physics. We consider Landau-Fermi (non-interacting electron) liquids,
in which case we show how to derive the Luther-Haldane
bosonization formulae from our methods, insulators,
incompressible Hall fluids (Laughlin fluids), and massless London
superconductors.
We then discuss duality, in the sense of eq. (1.10), for
two-dimensional systems and conclude with an analysis of
Laughlin fluids.

In Sect. 7, we consider perturbations of the systems discussed
in Sect. 6 by repulsive Coulomb(-Amp\`ere) two-body interactions,
along the lines sketched in eqs. (1.11) through (1.14).
We find the exact expression for the plasmon gap in a
metal, recover the consequences of the Anderson-Higgs
mechanism in a precise form and discuss the ``orthogonality
catastrophe" for static sources.

\vskip 0.5truecm

\underbar{Acknowledgements}. Much of the research reported
in this  paper was done when one of us (J.F.) was on a sabbatical leave from
E.T.H. at I.H.E.S., (March - June, 1993).
He thanks M. Berger and D. Ruelle for the kind hospitality
at I.H.E.S. and I. Affleck and Y. Avron for their interest in our work.
One of us (P.A. M.) thanks the E.T.H. for the kind hospitality
and A. Bassetto, F. Toigo for useful discussions.

{\bf 2.\ Preliminaries: gauge forms}

\vskip 0.5truecm

In this section we review some basic definitions and properties of
forms [9] and introduce the notion of gauge forms.

Let $M$ (the ``base space") be some  ($d$+1)-dimensional orientable
Riemannian manifold, to be identified with euclidean space-time. For
simplicity we assume the metric on $M$ to be flat. Typically, $M$ will be
an open subset of ${\bf R}^{d+1}$ or of a flat torus. Points in $M$ are
denoted by $x=(x^0, {\bf x})$, where $x^0$ denotes the euclidean time
coordinate of $x$.

Given an antisymmetric tensor field of rank $k$ on $M$, one defines
the associated differential form of rank $k$, or simply $k$-form, by
setting

$$
a^{(k)} (x) = {1\over k!} a_{\mu_1...\mu_k} (x)
dx^{\mu_1} \wedge...\wedge dx^{\mu_k}\eqno(2.1)
$$

where $\wedge$ is the wedge (antisymmetric tensor) product.
[We shall define all basic geometrical objects needed in our
analysis locally, in coordinate charts.]
The space of $k$-forms is a group, $\Lambda^k(M)$, under the
operation of  pointwise addition.

The exterior differential, $d$, mapping $k$-into $(k+1)$-forms, is
defined by

$$
da^{(k)} (x) = {1\over k!}
\partial_\mu a_{\mu_1...\mu_k} (x) dx^\mu \wedge
dx^{\mu_1} \wedge...\wedge dx^{\mu_k}. \eqno(2.2)
$$

One easily checks the key property

$$
dd = 0. \eqno(2.3)
$$

The $k$-th (de--Rham) cohomology group of $M$, $H^k_{deR}(M)$,
is the quotient group obtained by dividing the group of $k$-forms in the
kernel of $d$ (closed $k$-forms) by the image of $\Lambda^{k-1}(M)$
under $d$, (exact $k$-forms). Hence if $H^k_{deR}(M)=0$, the equation

$$
da^{(k)}=0
$$

has a solution of the form

$$
a^{(k)} = da^{(k-1)},\eqno(2.4)
$$

where $a^{(k-1)}$ is a $(k-1)$-form.
An element of $H^k_{deR}(M)$ is called a cohomology class.

One defines the
Hodge star, mapping $k$-forms to $(d+1-k)$-forms, by setting

$$
^*a^{(k)} (x) = {1\over k!}
{1\over (d+1-k)!}
a^{\mu_1...\mu_k} (x)
\epsilon_{\mu_1...\mu_k...\mu_{d+1}}
dx^{\mu_{k+1}}\wedge....\wedge dx^{\mu_{d+1}}. \eqno(2.5)
$$

The codifferential, $\delta$, mapping $k$-into $(k-1)$-forms, is defined
by

$$
\delta = ^* d^*(-1)^{(d+1)(k+1)+1}.\eqno(2.6)
$$

One defines an inner product between $k$-forms by setting

$$
(a^{(k)}, b^{(k)}) = \int d^{d+1} x a_{\mu_1... \mu_k} (x)
b^{\mu_1... \mu_k} (x) = \int a^{(k)} \wedge^* b^{(k)}. \eqno(2.7)
$$

Notice that

$$
(a^{(k)}, db^{(k-1)}) = (\delta a^{(k)}, b^{(k-1)}). \eqno(2.8)
$$

The laplacian on $k$-forms is given by

$$
\Delta = \delta d + d \delta \eqno(2.9)
$$

and a form $a^{(k)}$ satisfying

$$
\Delta a^{(k)} = 0 \eqno(2.10)
$$

is called harmonic; it is closed and is the canonical
representative of its cohomology class.

Next, we introduce the notion of gauge forms of
rank $k$, or generalized $U(1)$-connections of rank $k$ [7,8].
For $k$=1, it coincides with the standard notion of a
$U(1)$-gauge field or, in mathematical language, of
a $U(1)$-connection.

The natural framework to discuss gauge forms is the
theory of fibre bundles, but, in our definition,
we avoid that theory, postponing a more mathematical
discussion to a later remark.

\vskip 0.5truecm
\underbar{Definition}
\vskip 0.3truecm
Let ${\cal U} = \{U_i \}_{i \in I}$ denote a covering of
$M$ by open subsets with the property that the intersection
of any finite number of $U_i$'s is a contractible set.

Then a gauge form of rank $k, \tilde a^{(k)}$, is a collection
of $k$-forms $\{a^{(k)}_i \}_{i \in I}$ on ${\cal U}$ such
that

1) for $x \in U_i \cap U_j$

$$
a^{(k)}_i (x) - a^{(k)}_j (x) = d \lambda^{(k-1)}_{ij} (x),
\eqno(2.11)
$$

with $\lambda^{(k-1)}_{ij} \in \Lambda^{k-1} (U_i \cap U_j)$.

{}From (2.11) and (2.3) it follows that

$$
da^{(k)}_i (x) = da_j^{(k)} (x), \eqno(2.12)
$$

for $x$ in $U_i \cap U_j$ and hence $\{d a_i^{(k)} \}_{i \in I}$
defines a closed $(k+1)$- form, $f^{(k+1)} (\tilde a)$, on
$M$.

2) For every closed $(k+1)$-dimensional surface, $S_{(k+1)}$, in $M$,
$f^{(k+1)} (\tilde a)$ has the property that

$$
{1 \over 2\pi} \int_{S_{(k+1)}} f^{(k+1)} (\tilde a) \in {\bf Z}.
\eqno(2.13)
$$

The form $f^{(k+1)} (\tilde a)$ is called the curvature (or field
strenght) of $\tilde a^{(k)}$.

The space of gauge forms of rank $k$ is denoted by ${\cal A}^k$.
\vskip 0.5truecm

\underbar{Remark 2.1}. From the definition it follows
that a gauge form of rank $k>1$ can be seen as a generalization
of a $U(1)$-connection
(called a generalized $U(1)$--connection of rank $k$) on a
principal bundle, ${\cal P}^k$, whose transition functions
are given by the $(k-1)$-forms $\{\lambda^{(k-1)}_{ij} \}$.
This bundle is called a $U(1)$-fiber bundle of rank $k$, and
${\cal A}^k \equiv {\cal A}^k ({\cal P}^k)$ is the space of
$U(1)$-connections of
rank $k$ on ${\cal P}^k$. By equation (2.11), the difference of two
connections, $\tilde a^{(k)}, \tilde a'^{(k)} \in {\cal A}^k ({\cal P}^k)$, is
a globally defined $k$-form, i.e., $\tilde a^{(k)} - \tilde a'^
{(k)} \in \Lambda^k (M)$.
Hence, ${\cal A}^k$ is an affine space modeled on $\Lambda^k (M)$
i.e.,

$$
{\cal A}^k = \tilde a^{(k)} + \Lambda^k (M), \qquad \tilde a^{(k)}
\in {\cal A}^{k}.\eqno(2.14)
$$

Two $U(1)$-fiber bundles of rank $k,{\cal P}^k, {\cal P'}^k$
characterized by the transition functions $\{\lambda_{ij} \},
\{\lambda'_{ij}\}$, are said to be isomorphic if

$$
\lambda'^{(k-1)}_{ij} = \lambda^{(k-1)}_{ij} +
\lambda^{(k-1)}_i - \lambda_j^{(k-1)} + \zeta^{(k-1)}_{ij}
\eqno(2.15)
$$

for $\lambda^{(k-1)}_i \in \Lambda^{k-1} (U_i), \lambda^{(k-1)}_j
\in \Lambda^{k-1} (U_j)$, and $\zeta_{ij}^{(k-1)} \in \Lambda^{k-1}(U_i
\cap U_j)$
is a closed form with integral periods (i.e., its integral over an
arbitrary closed $(k-1)$ surface contained in $U_i \cap U_j$ is
an integer). An isomorphism class of $U(1)$-fiber bundles of rank $k$ is
called a $U(1)$-bundle of rank $k$. If $M$ is $k$-connected, the
$U(1)$-bundles of rank $k$ are classified by $H^{k+1} (M, {\bf Z})$, which is
isomorphic to the subgroup of $H^{k+1}_{deR} (M)$ given by the
cohomology classes of $(k+1)$-forms of integral periods. The classification
map associates to a bundle the cohomology class of ${1 \over 2\pi}
f^{(k+1)} (\tilde a)$, where $\tilde a^{(k)}$ is a connection on the
bundle [8].

\vskip 0.5truecm

The space of gauge forms, ${\cal A}^{k}$, carries an action of the
gauge group ${\cal G}^{k-1}$, whose elements, $\tilde \lambda^{(k-1)}$,
are collections $\{\lambda_i^{(k-1)}\}_{i \in I}$ of ($k$-1)-forms, with
the following patching property

$$
\lambda_i^{(k-1)} (x) - \lambda_j^{(k-1)} (x) = d \zeta^{(k-2)}_{ij}
(x), \eqno(2.16)
$$

for $x \in U_i \cap U_j$, with $\zeta_{ij}^{(k-2)} \in \Lambda^{k-2}$
$(U_i \cap U_j)$; (we set $\Lambda^{-1} (M) \equiv \{0\}$).

The action of ${\cal G}^{k-1}$ on ${\cal A}^{k}$ is given by

$$
a^{(k)}_i \longrightarrow a_i^{(k)} + d \lambda_1^{(k-1)}. \eqno(2.17)
$$

Notice that if $H^k_{deR} (M)=0$, then

$$
{\cal G}^{k-1} \simeq \Lambda^{k-1} (M) \eqno(2.18)
$$

i.e., $\tilde \lambda^{(k-1)}$ is determined by a globally
defined ($k$-1)-form, $\lambda^{(k-1)}$. Furthermore if
$H^{k+1}_{deR} (M)=0$, or, more generally, if the
cohomology class of $f^{k+1}(\tilde a)$ is zero, then

$$
{\cal A}^{k} \simeq \Lambda^k (M), \eqno(2.19)
$$

i.e., one can view $\tilde a^{(k)}$ as a globally
defined $k$-form $a^{(k)}$.

\vskip 0.5truecm

\underbar{Remark 2.2}.
Path--integrals over gauge forms (introduced later) can be defined
rigorously only on a distributional completion of ${\cal A}^{k}$.
A more careful discussion of the subject can be found in [10], for
$k$=1.

\vfill\eject

{\bf 3. \ Bosonization}

\vskip 0.5truecm

In this section we present the details of our method of
bosonization.

In the euclidean path-integral formalism, fermions of spin $S$
are described in terms of Grassmann fields $\Psi_\alpha, \Psi^*_\alpha,
\alpha=1$, ..., $2S+1$.

We consider a system of fermions whose euclidean  action, $I
(\Psi, \Psi^*)$, is local in $\Psi, \Psi^*$ and their derivatives
and invariant under the global $U(1)$-gauge tranformations

$$\eqalign{
\Psi_\alpha (x) & \longrightarrow e^{i \Lambda} \Psi_\alpha (x) \cr
\Psi^*_\alpha (x) & \longrightarrow e^{-i \Lambda} \Psi_\alpha^*
(x), \qquad\qquad\Lambda \in {\bf R}. \cr} \eqno (3.1)
$$

We couple the system to a $U(1)$-gauge field, $A_\mu$, by replacing
derivatives by covariant derivatives, thus gauging the symmetry
(3.1). Let $I(\Psi, \Psi^*, A)$ denote the  corresponding
gauge-invariant action. We define the effective action, $S(A)$,
of the system by setting

$$
e^{-S(A)} \equiv \int {\cal D} \Psi {\cal D} \Psi^*
e^{-I(\Psi, \Psi^*, A)}. \eqno(3.2)
$$

Let $j^{(1)}$ be a 1-form.
The Fourier transform of (3.2) is given by the following equation:

$$
e^{-\tilde S(j^{(1)})} \equiv \int {\cal D} A  e^{-S(A)} e^{i(A^{(1)},
j^{(1)})} \eqno(3.3)
$$

where ($\cdot$,$\cdot$) denotes the inner product between
forms defined in (2.7).

[The spin index, $\alpha$, of $\Psi_\alpha, \Psi^*_\alpha$ and
the index denoting the rank of the forms $A^{(1)}, j^{(1)},
\alpha^{(k)}$, etc... will be omitted in the following,
whenever there is no danger of confusion.]

Using the invariance of $I(\Psi, \Psi^*, A)$ under the
gauge transformation

$$
A \rightarrow A - d \Lambda \eqno(3.4)
$$

one can integrate the r.h.s. of (3.3) over all gauge transformations
(3.4). Using (2.8), we then obtain the constraint

$$
\delta j = ^*d^* j = 0, \eqno(3.5)
$$

i.e., a continuity equation for $j$. In this section we choose $M =
{\bf R}^{(d+1)}$, so that $H^k_{deR}(M)=0$ for all $k$.
[We define $\Lambda^k(M)$ to consist of $k$-forms
vanishing at infinity for all $k\geq 0$].

Then, according to (2.4), equation (3.5) can be explicitly
solved by introducing a $(d-1)$-form  $b \equiv b^{(d-1)}$
with

$$
j = {1 \over 2 \pi}\quad ^*db, \eqno(3.6)
$$

where the factor ${1 \over 2\pi}$ has been inserted for later
convenience. For $d>1$, the solution, $b$, of (3.6) is not unique, since
two forms, $b$, $b'$, differing by a term $d \lambda^{(d-2)}$ yields the
same $j$, thanks to property (2.3).
As a consequence, the action $S(j) \equiv S(db)$ of eq. (3.3)
is invariant under the gauge transformation

$$
b \rightarrow b + d \lambda^{(d-2)} \eqno(3.7)
$$

This is the action of the gauge group ${\cal G}^{d-2}$,
as defined in (2.16), on the space ${\cal A}^{d-1}$, with $\lambda^{(d-1)}$
globally defined because $H^{d-1}_{deR} (M)=0$.
Thus $b$ should be viewed as a gauge form of rank $d$-1 which
is globally defined because $H^d_{deR} (M)=0$.

[For later purposes, we use a notation, $\tilde S (db)$, slightly
different from the one adopted in the introduction,i.e. $\tilde S (b)$.]
\vskip 0.3truecm

\underbar{Remark 3.1}. If we consider a manifold $M$ with
$H_{deR}^{d-1} (M) \not = 0$ then two forms, $b, b'$,
differing by $d \lambda^{(d-2)}_i$ on the open set $U_i, i \in
I$, yield the same current $j$ if $\{\lambda_i^{(d-2)}\},i \in I$
satisfies property (2.16). In fact, locally,

$$
d(b - b')= dd \lambda_i^{(d-2)} = 0
$$

and, by property (2.16), $\{d \lambda_i^{(d-2)}\}$ defines a
global (closed) form on $M$. Also, in this situation, the
invariance group is exactly ${\cal G}^{d-2}$.

\vskip 0.3truecm

In terms of the ($d$-1)-form $b$, we may rewrite (3.3) as

$$
e^{-\tilde S(db)} \equiv \int {\cal D} [A] e^{-S(A)} e^{{1 \over 2\pi}
\int A \wedge db} \eqno(3.8)
$$

\noindent
where ${\cal D} [A] e^{- S(A)} (\cdot)$ denotes the measure
induced by ${\cal D} A e^{- S(A)}$ on the space of gauge orbits

$$
[A] = \{A':  A - A'= d \Lambda \}. \eqno(3.9)
$$

Next, we wish to prove a set of bosonization identities.
Let ${\cal D} [b] e^{-\tilde S{(db)}}$ denote the measure on the
gauge equivalence classes

$$
[b] = \{b' : b' = b + d \lambda^{(d-2)} \}. \eqno(3.10)
$$

Formally,

(1)
$$
\Xi \equiv \int {\cal D} \Psi {\cal D} \Psi^* e^{-I(\Psi, \Psi^*)}
= \int {\cal D} [b] e^{-\tilde S(db)}. \eqno(3.11)
$$

Let

$$
{\cal J}^\mu (\Psi, \Psi^*, A; x) = -i {\delta \over \delta A_\mu(x)}
I(\Psi, \Psi^*, A) \eqno(3.12)
$$

\noindent
so that ${\cal J}^\mu (\Psi, \Psi^*, 0) \equiv {\cal J}^\mu (\Psi,
\Psi^*)$ is the $U(1)$-current of the fermion system corresponding
to the global symmetry (3.1).

Then, for non coinciding points $\{x_1,..., x_n\}$, we have
that

(2)
$$\eqalign{
\langle{\cal J}^{\mu_1} (\Psi, \Psi^*; x_1) &...{\cal J}^{\mu_n} (\Psi,\Psi^*;
x_n)\rangle \cr
\equiv {1 \over \Xi} \int {\cal D} \Psi {\cal D} \Psi^*
& e^{-I(\Psi, \Psi^*)} \prod^n_{i=1} {\cal J}^{\mu_i} (\Psi, \Psi^*; x_i) \cr
= {1 \over \Xi} \int {\cal D} [b] & e^{-\tilde S(db)} \prod^n_{i=1}
{1 \over 2\pi} (^*db)^{\mu_i} (x_i) \equiv \langle \prod^n_{i=1}
{1 \over 2 \pi}
(^*db)^{\mu_i} (x_i) \rangle \cr} \eqno(3.13)
$$

\underbar{Proof}: Assuming one can interchange the order of integration
we have, formally, that

(1)
$$\eqalign{
& \int {\cal D} [b]  e^{-\tilde S(db)} = \int {\cal D} [b] \int {\cal D}[A]
\int {\cal D} \Psi, {\cal D} \Psi^* e^{-I(\Psi, \Psi^*, A)}
 e^{{i \over 2\pi} \int A \wedge db} = \cr
& \int {\cal D} [A] \int {\cal D}
\Psi  {\cal D} \Psi^*  e^{-I(\Psi, \Psi^*, A)} \delta (dA) =
\int {\cal D} \Psi {\cal D} \Psi^*  e^{-I(\Psi, \Psi^*)} \cr}
\eqno(3.14)
$$

By integrating by parts, we obtain that

(2)
$$\eqalign{
\langle\prod^n_{\ell=1} & {1 \over 2 \pi} (^*db)^{\mu_\ell} (x_\ell)\rangle=
\cr
\Xi^{-1} & \int {\cal D} [b] \int {\cal D} [A] \int {\cal D} \Psi
{\cal D} \Psi^* e^{-I(\Psi, \Psi^*, A)} \prod^n_{\ell=1} (- i {\delta \over
\delta A_{\mu_\ell} (x_\ell)}) e^{{i \over 2 \pi} \int A \wedge db} \cr
= \Xi^{-1} & \int {\cal D} [b] {\cal D} [A] \prod^n_{\ell=1}
\Bigl(i {\delta \over \delta A_\mu (x_\ell)} \Bigr) \Bigl(\int {\cal D} \Psi
{\cal D} \Psi^* e^{-I(\Psi, \Psi^*, A)} \Bigr)
e^{{i \over 2\pi}
\int A \wedge db} \cr
= \Xi^{-1} & \int {\cal D} [b] {\cal D} [A] \int {\cal D} \Psi
{\cal D} \Psi^* e^{-I(\Psi, \Psi^*, A)} \prod^n_{i=1} \Bigl(-i {\delta
\over \delta A_\mu (x_\ell)} \Bigr) I(\Psi, \Psi^*, A) \cr
& e^{{i \over 2\pi}
\int A \wedge db}
= \Xi^{-1} \int {\cal D} [A] \int {\cal D} \Psi {\cal D} \Psi^*
e^{-I(\Psi, \Psi^* A)} \delta (dA)
\prod^n_{\ell=1} {\cal J}^{\mu_\ell}
(\Psi, \Psi^*, A;  x_\ell). \cr} \eqno(3.15)
$$

Our explicit calculation, eq. (3.15), shows that identity
(3.13) also holds at coinciding points if the current correlation
functions in the fermion theory are defined by

$$\eqalign{
\langle \prod^n_{\ell = 1} & {\cal J}^{\mu_\ell}(\Psi, \Psi^*; x_\ell)
\rangle\equiv \cr
& \int {\cal D} [A] \delta (dA) \prod^n_{\ell=1} \Bigl(i {\delta
\over \delta A_{\mu_\ell} (x_\ell)} \Bigr) \int {\cal D} \Psi {\cal D} \Psi^*
e^{-I(\Psi, \Psi^*,A)}. \cr} \eqno(3.16)
$$

\vskip 0.5truecm

\underbar{Remark 3.2}. \ With the definition (3.16) the
current correlation functions are transversal in the fermionic theory:
by gauge invariance of $S(A)$, we have that for an arbitrary test
function $\chi$,

$$\eqalign{
\int {\cal D} [A] \delta (dA) \prod^{n-1}_{\ell=1} \Bigl(-i{\delta \over
\delta A_{\mu_\ell} (x_\ell)}\Bigr) & e^{-S(A + \alpha d \chi)} \cr
= \int {\cal D} [A] \delta (dA) \prod^{n-1}_{\ell=1} \Bigl(-i{\delta \over
\delta A_{\mu_\ell} (x_\ell)} \Bigr) & e^{-S (A)}, \cr} \eqno(3.17)
$$

and, differentiating (3.17) with respect to $\alpha$ and
setting $\alpha$ to $0$, we conclude that

$$\eqalign{
& \int d^{d+1} x_n \partial_{\mu_n} \chi(x_n) \int {\cal D} [A] \delta (dA)
\prod^n_{\ell=1} \Bigl(-i{\delta \over \delta A_{\mu_\ell} (x_\ell)} \Bigr)
e^{- S (A)} \cr
& = \int d^{d+1} x_n \partial_{\mu_n} \chi (x_n) \langle {\cal J}^{\mu_1}
(\Psi,
\Psi^*; x_1)...{\cal J}^{\mu_n}(\Psi, \Psi^*; x_n)\rangle = 0 \cr} \eqno(3.18)
$$

Note that if $I(\Psi^*, \Psi, A)$ is linear in e.g. the component $A_0$
then the definition of ${\cal J}^0$ through eq. (3.16) coincides with the one
given in eq. (3.12).

\vskip 0.5truecm

\underbar{Remark 3.3}. \ Instead of using measures on gauge equivalence
classes for $A$ and $b$ in eqs. (3.8), (3.11), (3.13).., one can add to
the actions $S (A)$ and $\tilde S(db)$, for $d > 1$, the standard
gauge fixing and Faddeev-Popov terms. For gauge forms of rank $k>1$, this
involves a tower of gauge fixings and ghosts, as explained e.g. in [11].

\vskip 0.3truecm

To summarize, equation (3.11) expresses the partition function of
our theory, originally formulated in terms of the fermionic field
$\Psi$, as the partition function of the bosonic gauge form $b$.

Equation (3.13) proves that the correlation functions of the
physical current of the fermionic theory, ${\cal J}^\mu (\Psi, \Psi^*)$,
are given by the correlation function of the dual of the curvature,
or ``field strength", $(^*db)^\mu$, of the gauge form $b$.
This result has some interesting applications. We consider a system
with an action $I_{tot} (\Psi, \Psi^*)$ given by a perturbation
of $I(\Psi, \Psi^*)$ by a current-current interaction, in the form
of a term, $I_{pert} (\Psi, \Psi^*)$, given by

$$\eqalign{
I_{pert} & (\Psi, \Psi^*) ={1 \over 2}  \int V_{\mu\nu} (x, y) {\cal J}^\mu
(\Psi, \Psi^*; x)
{\cal J}^\nu (\Psi, \Psi^*; y) \cr
\equiv & {1 \over 2} \Bigl( {\cal J} (\Psi, \Psi^*), V {\cal J} (\Psi,
\Psi^*) \Bigr) . \cr} \eqno(3.19)
$$

This perturbation, quartic in the $\Psi$ field, becomes quadratic
in the $b$ field. In fact, adopting definition (3.16), we obtain, for
the partition function $\Xi (V)$ of the perturbed theory, the expression

$$\eqalign{
\Xi (V) & = \int {\cal D} \Psi {\cal D} \Psi^* e^{-[I(\Psi, \Psi^*)
+ {1 \over 2} ({\cal J} (\Psi, \Psi^*), V {\cal J} (\Psi, \Psi^*))]} +\cr
&  \int {\cal D} [b] \int {\cal D} [A] \int {\cal D} \Psi {\cal D} \Psi^*
\Bigl(e^{{1 \over 2}({\delta \over \delta A}, V {\delta \over \delta
A})} e^{- I(\Psi, \Psi^*, A)} \Bigr) e^{{i \over 2\pi} \int A \wedge db} \cr
= & \int {\cal D} [b] \int {\cal D} [A] \Bigl( e^{{1 \over 2}({\delta
\over \delta A}, V {\delta \over \delta A})}
e^{{i \over 2 \pi} \int A \wedge db} \Bigr) e^{-S}
(A) \cr
= & \int {\cal D} [b] \int {\cal D} [A] e^{-{1 \over 8 \pi^2} (^*db, V^* db)}
e^{-S(A)}
e^{{i \over 2\pi} \int A \wedge db} \cr
= & \int {\cal D} [b] e^{-[\tilde S(db)+ {1 \over 8 \pi^2} (^*db, V^* db)]}.
\cr} \eqno(3.20)
$$

Clearly, these somewhat abstract identities become useful only if $\tilde
S(db)$
has a tractable form. In particular, if $S(A)$ is quadratic,
the perturbed theory has an action that is still quadratic in $b$.
In later sections, we briefly discuss some systems where this appealing
situation is encountered, i.e., where $S(A)$ is quadratic in A:
massless relativistic fermions in one dimension, the ``scaling
limit" of finite--density,
non--relativistic free fermions in arbitrary dimensions,
the ``scaling limit" of the bulk effective action of an
incompressible quantum hall fluid, or an insulator.

\vskip 0.5truecm
{\bf 4. \ Disorder fields and fermion correlation functions}
\vskip 0.5truecm

In the previous section we have reformulated a fermionic theory
with global $U(1)$-gauge invariance as a gauge theory
(for $d>1)$ of gauge forms of rank $d-1$.

Here we wish to discuss the question how correlation functions
of the field $\Psi$ and $\Psi^*$ can be expressed in the bosonic
theory.
A general result is that they involve a disorder field conjugate to the
gauge form $b$. We first propose a definition of disorder fields for
gauge forms in general terms.  A geometrical interpretation of our
definition, using the language of fiber bundles, is then provided later.

Let us denote by $\tilde S(db)$ the action of a gauge theory of ($d-1$)
forms in $M= {\bf R}^{d+1}$.
We choose $n$ points ${\underline x}$ = $\{x_1, ... ,x_n \}$
in ${\bf R}^{d+1}$ and define

$$
M_{\underline x} = {\bf R}^{d+1} \setminus \{x_1,..., x_n\}.\eqno(4.1)
$$

The cohomology group $H^{d-1}_{deR} (M_{{\underline x}})$
is trivial and
hence ${\cal G}^{d-2} \simeq \Lambda^{d-2}(M_{{\underline x}})$. However,
$H^{d}_{deR} (M_{{\underline x}}) \not = 0$; hence gauge
forms $\tilde b$ whose curvature belongs to a non-trivial
cohomology class in $H^{d}_{deR} (M_{\underline x})$ are not globally
defined on $M_{\underline x}$.

We choose $n$ non-zero integers, ${\underline q}$ =$\{q_1,...,
q_n\}$, satisfying $\sum^n_{i=1} q_i = 0$ and define the closed
$d$-form

$$
\varphi_{{\underline x}; {\underline q}} = \sum^n_{i=1} \varphi_
{x_i,q_i},\ {\rm where}\qquad \varphi_{x_i; q_i} = 2 \pi q_i \ ^*d \Delta^
{-1} \delta_{x_i}, \eqno(4.2)
$$

with $\delta_{x_i} \equiv \delta (x_i - z)$. One easily checks
that, for every closed $d$-dimensional surface $S_{(d)}$ in
$M_{\underline x}$,

$$
{1 \over 2 \pi} \int_{S_{(d)}} \varphi_{{\underline x}; {\underline q}}
\in {\bf Z}. \eqno(4.3)
$$

More precisely, for a $(d+1)$-dimensional ball, $B^i$, in ${\bf R}^{d+1}$
with $x_i \in B^i$, $x_j \not \in B^i$, for $j \not = i$, we have that

$$
{1 \over 2 \pi} \int_{\partial B^i} \varphi_{{\underline x};
{\underline q}} = q_i, \eqno(4.4)
$$

where $\partial$ denotes the boundary.
Thanks to equation (4.3), $\varphi_{{\underline x}; {\underline q}}$
is the curvature of a gauge form $\tilde \alpha_{{\underline x};
{\underline q}},$ of rank $d$-1 in $M_{\underline x}$.
Note that

$$
\delta \varphi_{{\underline x}; {\underline q}} =0 \eqno(4.5)
$$

and

$$
d \varphi_{{\underline x}; {\underline q}} = 2\pi \sum^n_{i=1} q_i \
^*\delta_{x_i}. \eqno(4.6)
$$

By (2.9), equations (4.5) and (4.6) imply that $\varphi_{{\underline x};
{\underline q}}$ is a harmonic form in $M_{\underline x}$.

In physics terminology, $\varphi_{x;q}$ is the vector potential
of a magnetic vortex at the point $x$ when $d$=1, while, for $d$=2, it is
the magnetic field of a monopole located at $x$, with magnetic charge
$q$.

The expectation value of a disorder field

$$
D({\underline x}, {\underline q}) \equiv \prod^n_{i=1} D(x_i, q_i)
\eqno(4.7)
$$

is given by

$$
\langle D ({\underline x}, {\underline q})\rangle = \{\Xi^{-1} \int {\cal D}
[b] e^{-\tilde S(db + \varphi_{{\underline x}; {\underline q}})} \}_{ren}
\eqno(4.8)
$$

where

$$
\Xi = \int {\cal D} [b] e^{-\tilde S(db)}
$$

is the partition function [12,13].

On the r.h.s. of (4.8)  a multiplicative renormalization
is usually necessary, but we shall often omit the subscript
``$ren$".
Formally, (4.8) can be interpreted as the expectation value of

$$
D({\underline x}, {\underline q}) = e^{-[\tilde S(db + \varphi_{{\underline x};
{\underline q}})- \tilde S(db)]} \eqno(4.9)
$$

Correlation functions involving $D({\underline x},{\underline q})$
and gauge--invariant functionals of $db$ and $b$, ${\cal F} (db, b)$,
are defined by

$$\eqalign{
\langle & D({\underline x},{\underline q}) {\cal F} (db,b)\rangle =  \cr
= & \Xi^{-1} \int {\cal D} [b] {\cal F} (db + \varphi_{{\underline x};
{\underline q}}, b + \tilde \alpha_{{\underline x}; {\underline q}})
e^{- \tilde S(db + \varphi_{{\underline x};{\underline q}})},
\cr} \eqno(4.10)
$$

with $\varphi_{{\underline x}; {\underline q}} = d \tilde \alpha_
{{\underline x}; {\underline q}}$.

Equation (4.10) can be understood  by introducing an action from the
left of\hfill\break $D(x_i, q_i)$ on ${\cal F}(db, b)$:

$$
D(x_i, q_i) {\cal F} (db, b) \equiv {\cal F} (db + \varphi_{x_i; q_i},
b + \tilde \alpha_{x_i; q_i}) D (x_i, q_i) \eqno (4.11)
$$

\vskip 0.5truecm
\underbar{Remark 4.1}. \ Definitions (4.8) and (4.10) can
be rephrased in the language of $U(1)$-fiber bundles of rank $d-1$.
First of all, we note that $M_{\underline x}$ is $d-1$-connected and that

$$
H^d (M_{\underline x}, {\bf Z}) \simeq \underbrace{{\bf Z} \oplus....
\oplus {\bf Z}}_{\rm n - times}, \eqno(4.12)
$$

for ${\underline x}= (x_1,..., x_n)$.

{}From Remark 2.1 it follows that

1) $U(1)$-bundles of rank $d-1$ on $M_{\underline x}$ are characterized
by the (de Rham) cohomology class of the curvature of the
generalized $U(1)$-connection of rank $d-1$ on the bundle;

2) if the curvature is given by $\varphi_{{\underline x};{\underline q}}$,
eq. (4.4), $q_i$ is precisely the integer belonging to the $i^{th}$
{\bf Z} group on the l.h.s. of (4.12). The set of integers
${\underline q}$ therefore completely characterizes the bundle.

Let ${\cal A}^{d-1}_{{\underline x};{\underline q}}$ denote the space of
connections of a $U(1)$-fiber bundle of rank $d$-1 on $M_{\underline x}$
characterized by the integers ${\underline q}$. Then eq. (4.10)
can be rewritten as

$$
\langle D ({\underline x}, {\underline q}) {\cal F} (db, b)\rangle
= \Xi^{-1} \int_{{\cal A}^{d-1}_{{\underline x};{\underline q}}}
{\cal D} [\tilde b] e^{-\tilde S(f(\tilde b))} {\cal F} (f(\tilde b),
\tilde b) \eqno(4.13)
$$

where $f(\tilde b)$ is the curvature of a connection $\tilde b$
belonging to ${\cal A}^{d-1}_{{\underline x};{\underline q}}$ [10,13].
In fact, by (2.14), ${\cal A}^{d-1}_{{\underline x}; {\underline q}}$
is an affine space, and every connection $\tilde b^{(d-1)}$ can
be written as

$$
\tilde b^{(d-1)}= \tilde b^{(d-1)}_0 + b, \eqno(4.14)
$$

with

$$
f(\tilde b^{(d-1)}) = f (\tilde b_0^{(d-1)}) + db, \eqno(4.15)
$$

where $\tilde b^{(d-1)}_0$ is a reference connection. Eq. (4.10)
is obtained by choosing

$$
\tilde b^{(d-1)}_0 = \tilde \alpha_{{\underline x}; {\underline q}}.
\eqno(4.16)
$$

By gauge invariance expression (4.13) is independent of the
choice of the reference connection.

\vskip 0.5truecm
It is well known that from the euclidean correlation functions
of $db(x)$ (satisfying a suitable variant of Osterwalder-Schrader
axioms) [14] one can reconstruct the vacuum state $|0>$, the Hilbert space
${\cal H}$, the Hamiltonian $H$ and field operators $\hat j({\bf x})=
\hfill\break^*\widehat{db}({\bf x})$
of a quantum field theory. The relation between the euclidean
field $db$ and the field operator $\widehat{db}$ is given
as follows:
Let $x^0_1 < x^0_2...< x^0_n$, and define

$$
\hat {\cal O} (x) \equiv e^{- x^0 H} \hat {\cal O} ({\bf x})
e^{x^0 H}. \eqno(4.17)
$$

Then

$$
\langle db (x_1)...db (x_n)\rangle = \Xi^{-1} \int {\cal D} [b]
e^{- \tilde S(db)} db (x_1)... db(x_n) = \langle 0| \widehat{db} (x_1)...
\widehat{db} (x_n) |0\rangle. \eqno (4.18)
$$

Extending this result, it has been proved in [13],
that from the correlation functions of the disorder
fields (satisfying a suitable version of the O.S.
axioms) one can reconstruct soliton field operators
$\hat S_q ({\bf x})$ such that, for $x^0_1 <...< x^0_n,$

$$
\langle D ({\underline x}, {\underline y})\rangle
= \langle 0| \hat S_{q_1} (x_1)..
\hat S_{q_n} (x_n) |0\rangle \eqno(4.19)
$$

Similar identies hold for mixed correlation functions
involving disorder fields and gauge-invariant functionals
of $db$ and $b$.

{}From (4.17) and (4.19) it follows that one may consider
$D(x,q)$ as the euclidean field corresponding to the
soliton field operator $\hat S_q ({\bf x})$ at euclidean
time $x^0$.

Given a gauge form $\tilde b^{(d-1)}$ and a $d$-dimensional
surface $\Sigma_{(d)}$, one defines the ``Wilson operator
of rank $(d-1)$" by

$$
W_\alpha (\Sigma_{(d)}) =
\Bigl\{ e^{i \alpha \int_{\Sigma_{(d)}} f (\tilde b^{(d-1)})}
\Bigr\}_{ren} , \eqno (4.20)
$$

where $\alpha$ is a real number.
We observe that if $\tilde b^{(d-1)}$ is a globally defined
form, denoted by $b$, then

$$
W_\alpha (\Sigma_{(d)}) = \Bigl\{e^{i \alpha \int_{{\cal L}_{(d-1)}}
b}\Bigr\}_{ren}, \eqno(4.21)
$$

where ${\cal L}_{(d-1)}$ is the boundary, $\partial \Sigma_{(d)}$,
of $\Sigma_{(d)}$, and $W_\alpha (\Sigma_{(d)})$ coincides with the
ordinary Wilson loop when $d=2$.
On the r.h.s. of (4.20), (4.21) a multiplicative renormalization, depending
on $\alpha$ and ${\cal L}_{(d-1)}$ is usually necessary, but we shall often
omit the subscript $ren$.

If $\Sigma_{(d)}$ is contained in a fixed-time (hyper) plane we
shall write
$
W_\alpha ({\cal L} _{(d-1)})
$,
instead of $W_\alpha (\Sigma_{(d)})$, and we denote by $\hat W_\alpha
({\cal L}_{(d-1)})$ the corresponding field operator.
The soliton operator and the Wilson operator $\hat W_\alpha ({\cal L}_
{(d-1)})$ satisfy the ``dual algebra"[15, 5].

$$
\hat S_q ({\bf x}) \hat W_\alpha ({\cal L}_{(d-1)}) = \cases{e^{i 2\pi
\alpha q}  \hat W ({\cal L}_{(d-1)}) \hat S_q ({\bf x}), & {\rm for} $
{\bf x} \in {\rm int} \Sigma_{(d)}$ \cr
\hat W_\alpha ({\cal L}_{(d-1)}) \hat S_q ({\bf x}), & {\rm for} $ {\bf x}
\not \in {\rm int} \Sigma_{(d)} $} \eqno(4.22)
$$

By $T_\epsilon$ we denote translation by $\epsilon$ in the
positive time direction, and let $x = (0, {\bf x})$. Then
the dual algebra implies the following equation:

$$
{\lim\limits_{\epsilon \downarrow 0}}\langle ... D({\underline x},
{\underline q})...W_{- \alpha}
(T_{- \epsilon} {\cal L}_{(d-1)}) W_\alpha (T_\epsilon
{\cal L}_{(d-1)})...\rangle =
$$

$$
= {\lim\limits_{\epsilon \downarrow 0}}\langle... e^{- i \alpha \int_{T_{-
\epsilon} \Sigma_{(d)}} (db + \varphi_{{\underline x}; {\underline q}})}
e^{i \alpha \int_{T_\epsilon \Sigma_{(d)}} (db + \varphi_{{\underline x};
{\underline q}})} D ({\underline x}, {\underline q})...\rangle
$$
$$
={\lim\limits_{\epsilon \downarrow 0}} \  e^{i \alpha (\int_
{T_\epsilon \Sigma_{(d)}} \varphi_{{\underline x}; {\underline q}}
- \int_{T_{-\epsilon} \Sigma_{(d)}} \varphi_{{\underline x}; {\underline q}})}
\langle... e^{-i \alpha \int_{T_ {-\epsilon} \Sigma_{(d)}} db + i \alpha
\int_{T_\epsilon \Sigma_{(d)}} db} D ({\underline x}, {\underline
q})...\rangle
$$
$$
= \exp  \Bigl(i \alpha \sum_{j: {\bf x}_j \in {\rm int} \Sigma_{(d)}}
q_j\Bigr) \langle ...
D({\underline x}, {\underline q})...\rangle . \eqno(4.23)
$$

Let $\hat \Psi_\alpha$ and $\hat {\cal J}^\mu$ denote the fermion field
operator and the current operator reconstructed from the correlation
functions of $\Psi_\alpha , \Psi^*_\alpha$ and ${\cal J}^\mu (\Psi,
\Psi^*)$ in the fermionic theory and let $\hat {\cal Q}_{\Sigma_{(d)}}$
denote the
$U(1)$ charge operator associated with a $d$-dimensional surface,
$\Sigma_{(d)}$, contained
in the  time-zero plane; formally

$$
\hat {\cal Q}_{\Sigma_{(d)}} = \int_{\Sigma_{(d)}} d^d x  \hat {\cal J}^0
({\bf x}). \eqno(4.24)
$$

The fact that $\hat \Psi_\beta^\# ({\bf x}) = \hat \Psi_\beta
({\bf x}), \Bigl(\hat \Psi_\beta^\dagger ({\bf x}) \Bigr)$
carries charge $+1 (-1)$ localized at ${\bf x}$ is summarized in the equation

$$
\langle 0| T  \Bigl(... e^{-i \alpha \hat {\cal Q}_{\Sigma_{(d)}}}
\hat \Psi_\beta^\#
({\bf x}) e^{i \alpha \hat {\cal Q}_{\Sigma_{(d)}}}...\Bigr) |0\rangle
$$

$$
= {\lim\limits_{\epsilon \downarrow 0}} \langle 0| T \Bigl(...e^{-i \alpha
\hat {\cal Q}_{T_{-\epsilon} \Sigma_{(d)}}} \hat \Psi^\#_\beta ({\bf x})
e^{i \alpha \hat {\cal Q}_{T_\epsilon \Sigma_{(d)}}}... \Bigr) |0\rangle
$$

$$
= {\lim\limits_{\epsilon \downarrow 0}} \langle T \Bigl(...e^{-i \alpha
\int_{T_
{- \epsilon} \Sigma_{(d)}} d^d x {\cal J}^0 ({\bf x})} \Psi^\#_\beta
(0, {\bf x}) e^{i \alpha \int_{T_\epsilon \Sigma_{(d)}} d^d {\cal J}^0
({\bf x})} ... \Bigr) \rangle
$$

$$
= \langle T  \Bigl(...\Psi^\#_\beta (0, {\bf x})...\Bigr)\rangle
\cases {e^{(\pm) i
\alpha} & $ {\bf x} \in {\rm int} \Sigma_{(d)} $ \cr
1 & $ {\bf x} \not \in \Sigma_{(d)} $ }
$$

$$
= \langle 0|T  \Bigl(.... \hat \Psi^\#_\beta ({\bf x})... \Bigr) |0\rangle
\cases{
e^{(\pm) i \alpha} & $ {\bf x} \in {\rm int} \Sigma_{(d)} $ \cr
1 & $ {\bf x} \not \in \Sigma_{(d)}$ }, \eqno(4.25)
$$

where $T$ (..) denotes euclidean time ordering.

We recall that ${\cal J}^0 (\Psi, \Psi^*)$ is represented in the
bosonic theory by ${1 \over 2 \pi} (^*db)^0$. Comparison of
(4.23) and (4.25) then yields:

$$
\Psi_\alpha (x) \propto \ D(x, 1), \qquad \Psi^*_\alpha (x) \propto \ D
(x, -1), \eqno(4.26)
$$

or, in operator language,

$$
\hat \Psi_\alpha^* (x)  \propto  \hat S_1 (x), \qquad
\hat \Psi_\alpha (x)  \propto  \hat S_{-1} (x), \eqno(4.27)
$$

where $\propto$ means ``proportional to".
Hence, fermion fields in the bosonic theory are proportional
to disorder fields, and fermions can be viewed as solitons of the
bosonic theory.
The form of the proportionality (4.26) is made explicit
for relativistic models in one dimension in section 5.

There is a class of models where the relation between
disorder fields and fermion fields can be made more
explicit. Consider a system of fermions with an action
$I(\Psi, \Psi^*)$, perturbed by a current-current
interaction (3.19).
We denote by $\langle(\cdot)\rangle^V$ the euclidean expectation
value of the perturbed system and by $\langle(\cdot)\rangle_A$
the expectation value corresponding to the unperturbed
gauge-invariant action $I(\Psi, \Psi^*, A)$. Suppose
that the fermion correlation functions are expressible
through a formula

$$
\langle T \Bigl(\Psi^*_{\alpha_1} (x_1)...\Psi^*_{\alpha_n} (x_n)
\Psi_{\alpha_{n+1}}
(x_{n+1})...\Psi_{\alpha_{2n}} (x_{2n}) \Bigr)\rangle_A =
\int d \mu_{{\underline x};
{\underline \alpha}} (J^{(1)}) e^{i(A, J)} , \eqno(4.28)
$$

where $d\mu_{{\underline x};{\underline \alpha}} (J^{(1)})$ is a
measure on the space of 1-forms $J^{(1)}$. The transformation
property under $U(1)$-gauge transformations of the fermion fields,
implies that $d \mu (J^{(1)})$ is supported on forms satisfying.

$$
\delta J^{(1)} (z) = \sum^n_{i=1} \delta_{x_i} (z) -
\sum^{2n}_{i=n+1} \delta_{x_i} (z). \eqno(4.29)
$$

Let $q_i =1$ for $i=1, ..., n, q_i= -1$ for $i=n+1, ... ,2n$.
Then, with $\varphi_{{\underline x}; {\underline q}}$ as
in (4.2),

$$
\Psi^*_{\alpha_1} (x_1)...\Psi^*_{\alpha_n}(x_n) \Psi_{\alpha_{n+1}}
(x_{n+1}))....
\Psi_{\alpha_{2n}} (x_{2n})
e^ {{i \over 2\pi} \int A \wedge \varphi_{{\underline x}; {\underline q}}}
\eqno(4.30)
$$

is gauge-invariant. Following (3.20) we derive the identity

$$\eqalign{
\langle T & (\Psi^*_{\alpha_1} (x_1) ... \Psi_{\alpha_{2n}} (x_{2n}) )
\rangle^V =
\Xi  (V)^{-1} \int {\cal D} [b] \int {\cal D} [A] {\cal D} \Psi
{\cal D} \Psi^*
T (\Psi^*_{\alpha_1} (x_1)... \cr
\Psi_{\alpha_{2n}} & (x_{2n})) e^{{i \over 2 \pi} \int A \wedge \varphi
_{{\underline x}; {\underline q}}} e^{ - I (\Psi, \Psi^*, A)
- {1 \over 2}  ({\cal J} (\Psi, \Psi^*, A), V {\cal J} (\Psi, \Psi^*,A))}
e^{{i \over 2\pi} \int A \wedge db} \cr
= & \Xi (V)^{-1} \int {\cal D} [b] \Bigl(\int {\cal D} [A]
e^{- S(A)}
\langle T  (\Psi^*_{\alpha_1} (x_1)... \Psi_{\alpha_{2n}} (x_{2n}))
\rangle _A \cr
& e^{{i \over 2\pi} A \wedge (db + \varphi_{{\underline x}; {\underline
q}})} e^{-{1 \over
8\pi^2} (^*(db + \varphi_{{\underline x}; {\underline q}}),
V^*(db + \varphi_{{
\underline x}; {\underline q}}))} \Bigr) \cr} \eqno(4.31)
$$

Inserting (4.28) and integrating over $A$, we obtain that

$$\eqalign{
\langle T & (\Psi^*_{\alpha_1} (x_1) .... \Psi_{\alpha_{2n}}
(x_{2n}))\rangle^V = \cr
\Xi & (V)^{-1} \int {\cal D} [b] \int d\mu_{{\underline x};{\underline
\alpha}} (J) e^{-\tilde S(db + \varphi_{{\underline x}; {\underline q}}+ 2
\pi^*
J)}
e^{-{1 \over 8 \pi^2} (^*(db+ \varphi_{{\underline x}; {\underline q}}),
V^*(db + \varphi_{{\underline x}, {\underline q}}))} \cr} \eqno(4.32)
$$

The r.h.s. of (4.32) has the desired form of an expectation value of a
disorder field.

\vskip 0.3truecm

\underbar{Remark 4.2}. The representation in eq. (4.28) of correlation
functions of fermions, holds, e.g., for a system of free
non-relativistic fermions with chemical potential $\mu$, at
temperature $T$.
The measure in (4.28) is defined as follows:
let ${\underline \omega} = \{\omega_1,...,\omega_n \}$ denotes
a set of (brownian) paths joining ${\underline x}= \{x_1,...,x_n \}$
to ${\underline y} = \{x_{n+1},..., x_{2n} \}$, and let ${\underline\alpha}=
\{\alpha_1, ..., \alpha_n\}, {\underline \delta}= \{\alpha_{n+1},...,
\alpha_{2n} \}$.
Formally we set

$$
J_{\underline \omega} (z) = \sum^n_{\ell=1} \Bigl(d z^0 \delta
({\bf z} - \omega_\ell (z^0))
+dz^i {dz_i \over dz^0} \delta ({\bf z} - \omega_\ell (z^0))
\Bigr) . \eqno(4.33)
$$

Then

$$
\int  d \mu (J_{\underline \omega}) {(\cdot)}  = \sum_{\pi \in \Sigma_n}
(-1)^{\sigma(\pi)} \prod^n_{k=1} \Bigl(\sum_{\ell_k = 0,1,..} (-1)^{\ell_k}
$$
$$
\Theta  (y^0_{\pi(k)} + \ell_k \beta - x^0_k) \Bigr)
\int\limits_{\scriptstyle\omega(x^0_k)=
{\bf x}_k\atop\scriptstyle\omega(y^0_{\pi(k)}+\ell_k \beta) =
{\bf y}_\pi(k)}
\prod^n_{k=1} [{\cal D} \omega_k
$$
$$
\exp \Bigl\{-\Bigl[\int^{y^0_{\pi(k)}+ \ell_k \beta}_{x^0_k} dt {m \over 2}
\dot\omega^2_k (t)\Bigr\} e^{\mu(y^0_{\pi (k)} + \ell_k \beta - x^0_k)}
\delta_{\alpha_k, \delta_{\pi (k})}\Bigr] (\cdot) \eqno(4.34)
$$

where $\Sigma_n$ is the group of permutations of $n$ objects, $\sigma(\pi)$
is the signature of $\pi \in \Sigma_n$, $\Theta$ denotes the Heaviside step
function, $m$ the mass of the particles and $\beta = {1 \over kT}$, where
$k$ is the Boltzmann constant. For a derivation of this formula see
[16]. In [17] the factor $(-1)^{\ell_k}$ was erroneously missing.
A similar formula
for relativistic, massive, spin ${1 \over 2}$ fermions can be found in
[18] [19].

\vskip 0.5truecm

To conclude this section we recall an interesting relation
between Wilson loops and disorder fields, 't Hooft duality [5].
Let $\Sigma^R_{(d)}$ denote a $d$-dimensional ball of radius $R$ in the
euclidean time-zero plane, let ${\cal L}^R_{(d-1)} \equiv \partial
\Sigma^R_{(d)}$
and let $W_\alpha ({\cal L}^R_{(d-1)})$ be the corresponding Wilson loop of
rank $d$-1. Then \underbar{'t Hooft duality}
is the following conjecture, verified in many concrete models.
If, for any $c > 0$,$\alpha \not = 0$,

$$
e^{c | {\cal L}^ R_{(d-1)}|}  \langle  W_\alpha(\Sigma^R_{(d)}) \rangle
\rightarrow 0,\ {\rm as} \
R \rightarrow \infty, \eqno(4.35)
$$

where $|{\cal L}|$ denotes the volume of ${\cal L}$, then the
expectation value
$\langle D(x, 1; y, -1)\rangle$ of the disorder field
$D(x,1; y,-1)$, (with $x, y$ in the
time-zero plane) has decay slower than exponential in $(x-y)$.
If, for some $c < \infty$,

$$
\langle W_\alpha ({\cal L}^R_{(d-1)})\rangle
\geq e^{-c|{\cal L}^R_{(d-1)}|}
\ {\rm as}\ R \rightarrow \infty, \eqno(4.36)
$$

i.e., the Wilson loop has perimeter decay or slower, then either
$\langle D(x, 1; y-1)\rangle $ has exponential decay in $(x-y)$, or
the density, $^*(db)^0$, correlation functions have a gapless mode.

There is also a  criterion for the existence
of a global $U(1)$-charge.

In the euclidean formalism the charge operator
$\hat {\cal Q}$ for the bosonic theory is defined
by the following (weak) limit:

$$
e^{i \alpha 2\pi \hat {\cal Q}} = {\lim\limits_{R \rightarrow \infty}}
\ {\hat W_\alpha ({\cal L}^R_{(d-1)}) \over \langle 0 |\hat W_\alpha
({\cal L}^R_
{(d-1)}) |0\rangle }\ . \eqno(4.37)
$$

The denominator is needed in order to ensure that the charge
(if it exists) annihilates the vacuum $|0>$.
A heuristic criterion for the existence of the limit in (4.37) is
that $\langle 0| \hat W_\alpha ({\cal L}^R_{(d-1)})|0\rangle$
= $\langle W_\alpha ({\cal L}^
R_{(d-1)})\rangle $ have, at most, perimeter decay.
This criterion has been proved to be correct for many lattice gauge
theories [20]. We propose to extended this criterion, as well  on
't Hooft duality, to systems of condensed matter physics in the
bosonized representation developed in this and the last section.

\vskip 0.5truecm
{\bf 5. \ Relativistic massless fermions in $d$=1}
\vskip 0.5truecm

In this section we show that the construction outlined in
sects. 3 and 4 reduces to ordinary abelian bosonization [21]
if $I(\Psi, \Psi^*)$ is the action of massless free Dirac
fermions in $d$=1.
We start by recalling the main formulas of abelian bosonization.
Let $\gamma^i, i = 0, 1, 5$ be two--dimensional euclidean Dirac matrices,
identified, in the chiral basis, with the Pauli matrices $\sigma_2,
\sigma_1, \sigma_3$. The euclidean Dirac operator is given by

$$
\parz = \gamma^0 \partial_0 + \gamma^1 \partial_1.
$$

Setting $x = ix^0 + x^1, \bar x = - ix^0 + x^1, \partial \equiv
{\partial \over \partial x}, \bar \partial \equiv {\partial \over
\partial \bar x}$:

$$
\parz = 2 \left(\matrix{
0 & \bar\partial\cr
\partial & 0\cr}\right).
$$

The euclidean action of the massless Dirac field is given by

$$
I(\Psi, \Psi^*)= \int d^2 x \Psi^*_\alpha \parz \Psi_\alpha(x)
\eqno(5.1)
$$

where $\Psi_\alpha, \Psi^*_\alpha =1,2$ are two--component Grassmann
fields. The two--point function derived from (4.3) is given by

$$
\langle \Psi_\alpha (x) \Psi^*_\delta(y)\rangle
= i \parz \Delta^{-1} (x-y)=
$$

$$
=\cases{
{i\over 2\pi} (\bar x - \bar y)^{-1},\ & $\alpha=1, \delta = 2$,\cr
{i \over 2\pi} (x-y)^{-1}, \ & $\alpha =2, \delta=1$, \cr
0,\ & {\rm otherwise},  \cr} \eqno(5.2)
$$

where $\Delta$ is the two dimensional laplacian.

The euclidean action for the massless, relativistic, real scalar field,
$\Phi$, is given by

$$
\tilde S(d \Phi) = {1 \over 8 \pi} \int d^2 x (\partial_\mu \Phi)^2 (x) =
{1 \over 8 \pi} (d \Phi, d \Phi)
\eqno(5.3)
$$

One can characterize the expectation value corresponding to the
action (5.3) by

$$
\langle e^{i(\Phi,f)}\rangle =
\cases{
e^{- 2\pi(f,\Delta^{-1}f)}, & $\tilde f(0)=0$\cr
0 & $\tilde f(0)\not = 0$, \cr}\eqno(5.4)
$$

where $f\in {\cal S}({\bf R}^2)$ and $\tilde f$ denotes its Fourier
transform.
Formally one can define a
``two-point function" derived from (5.3), (5.4) given by

$$
``\langle \Phi(x) \Phi (y)\rangle = 4 \pi \Delta^{-1} (x - y) \equiv -2
{\rm ln} | x - y|". \eqno(5.5)
$$

The simplest bosonization formulas [...], proved by direct computation,
are the following: the euclidean correlation functions of
${\cal J}^\mu(\Psi,\Psi^*)=\ :\Psi^*\gamma^\mu\Psi:$\ and
$:\Psi^*{ (1\pm\gamma_5)\over 2}\Psi$:\ , normal ordered in the fermionic
theory, are identical to the euclidean correlations of
${1\over 2\pi}\epsilon_{\mu\nu}\partial^\nu\Phi$ and $:e^{\pm i\Phi}$:\ ,
normal ordered in ths bosonic theory. The normal ordering chosen in
the bosonic theory can be characterized formally by

$$
: e^{i\alpha\Phi(x)} : =
e^{i\alpha\Phi(x)} (2\pi)^{\alpha^2}
e^{2\pi\alpha^2\Delta^{-1}(x,x)}\eqno(5.6)
$$

[For a rigorous definitions one uses a point-splitting regularization].

One can exhibit also explicit bosonization formulas for the fields
$\Psi$ and $\Psi^*$. Formulas for the above bilinear expressions then
follow by taking limits [22]. We briefly recall some details
of the result.

Let ${\underline x} = \{x_1,..., x_n\}, y = \{y_1,...,y_n\},
{\underline q} = \{q_i\}^n_{i=1}, {\underline q}' =
\{q'_j \}^n_{j=1}, q_i = -1, q'_j = +1.
$.

One first constructs the corresponding closed 1--form  (4.2),
$\varphi_{{\underline x}, {\underline y}; {\underline q},
{\underline q'}} \equiv \varphi_{{\underline x}, {\underline y}}$
and the corresponding rank-0 gauge form $\tilde \alpha_{{\underline x},
{\underline y}; {\underline q}, {\underline q'}} \equiv \tilde \alpha_
{{\underline x}, {\underline y}}$. One can give an explicit expression
for $\tilde \alpha_{{\underline x}, {\underline y}}$ as the multivalued
function

$$
\tilde \alpha_{{\underline x}, {\underline y}} (z) = \sum^n_{i=1}
\Bigl(\arg (x_i - z) - \arg (y_i - z) \Bigr), \eqno(5.7)
$$

where $\arg\ x, x \in {\bf R}^2$ denotes the argument of
$ix^0 + x^1$.

According to definition (4.8) the expectation value of
the disorder field  is given by

$$
\langle D ({\underline x}, {\underline q}, {\underline y},
{\underline q'})\rangle \equiv \langle D ({\underline x}, {\underline
y})\rangle
= \Bigl\{{\int {\cal D} \Phi e^{-\tilde S(d \Phi + \varphi_{{\underline x},
{\underline y}})} \over \int {\cal D} \Phi e^{-\tilde S(d \Phi)}}\Bigr\}_{ren}.
\eqno(5.8)
$$

To give the explicit form of the renormalization, notice that
the term quadratic in $\varphi_{{\underline x}, {\underline y}}$
in $\tilde S(d \Phi + \varphi_{{\underline x},{\underline y}})$
is logarithmically divergent: formally

$$
{1 \over 8 \pi} (\varphi_{{\underline x}, {\underline y}},\varphi_
{{\underline x}, {\underline y}}) = {1 \over 8 \pi} \sum^n_{i,j=1}
\{2 q_i q'_j \Delta^{-1} (x_i, y_j)+ q_i q_j \Delta^{-1}(x_i, x_j)+
q'_i q'_j \Delta^{-1} (y_i, y_j)\}
\eqno(5.9)
$$

A multiplicative renormalization is then necessary, as mentioned
after equation
(4.8), to eliminate the terms with coinciding points in (5.9).
This can be done as follows: let $S_\delta (x)$ denote a small
ball of radius $\delta$ around $x$, and set

$$
c (\delta) \equiv {1 \over 8 \pi} \Bigl(\sum^n_{i=1} \int_{S_\delta (x_i)}
+ \sum^n_{j=i} \int_{S_\delta(y_j)} \Bigr) \varphi_{{\underline x},
{\underline y}} \wedge^* \varphi_{{\underline x}, {\underline y}}
$$

$$
{\mathop{\longrightarrow}\limits_{\delta \rightarrow 0}}\ {1 \over 8 \pi}
\Bigl(\sum^n_{i=1} \Delta^{-1}
(x_i, x_i) + \sum^n_{j=1} \Delta^{-1} (y_j, y_j) \Bigr).
\eqno(5.10)
$$

We define a regularized action by

$$
\tilde S^\delta (d \Phi + \varphi_{{\underline x}, {\underline y}}) =
{1 \over 8 \pi} \int_{{\bf R}^2 \backslash S_\delta
({\underline x}, {\underline y})}
(d \Phi + \varphi_{{\underline x}, {\underline y}}) \wedge^*
(d \Phi + \varphi_{{\underline x}, {\underline y}}) - c(\delta)
\eqno(5.11)
$$

where $S_\delta ({\underline x}, {\underline y}) =
\cup^n_{i=1} S_\delta (x_i) \cup^n_{j =1} S_\delta (y_j)$.

The precise definition of disorder field is given by

$$
\langle D ({\underline x}, {\underline y})\rangle
= {\lim\limits_{\delta \downarrow
0}}\ {\int {\cal D} \Phi e^{-\tilde S^\delta (d \Phi + \varphi_{{\underline x},
{\underline y}})} \over \int {\cal D}\Phi e^{-\tilde S(d \Phi)}} \eqno(5.12)
$$

[From (5.10) it follows that the r.h.s. of (5.12) can be written as the
r.h.s. of (5.8) times a multiplicative (infinite) renormalization.

$$
``e^{{1 \over 8 \pi} \Bigl(\sum^n_{i=1} \Delta^{-1} (x_i, x_1) + \sum^n_{j=1}
\Delta^{-1} (y_j, y_j)\Bigr)}"]. \eqno(5.13)
$$

With the disorder field defined as above, one can prove the bosonization
identity:

$$\eqalign{
\langle T & \Bigl(\Psi^*_{\alpha_1} (x_1)... \Psi^*_{\alpha_n}
\Psi_{\delta_1}(y_1)...
\Psi_{\delta_n} (y_n)\Bigr)\rangle \cr
= & (2 \pi)^{-{n \over 2}} \langle D ({\underline x},{\underline y})
\prod^n_{i=1}
: e^{-{1 \over 2}(-1)^{\alpha_i} \Phi(x_i)} :
\prod^n_{j=1} : e^{-{1 \over 2} (-1)^{\delta_j} \Phi(y_j)}:\rangle. \cr}
\eqno(5.14)
$$

Adding to $I(\Psi, \Psi^*)$ a current-current interaction (3.19), with
${\cal J} (\Psi, \Psi^*) =  :\Psi^* \gamma_\mu \Psi$:, is equivalent
to adding the term ${1 \over 4 \pi^2} (^*d \Phi, V^*d \Phi)$ to $\tilde S
(d \Phi)$
in the bosonic theory. If we denote by $\langle (\cdot) \rangle^V$ the
corresponding expectation value, (5.14) holds for the perturbed theory
replacing $\langle (\cdot) \rangle$ by $\langle (\cdot)\rangle^V$.

With the convention of writing the disorder field to the left of all
functionals of $\Phi$, the bosonization formula (5.14) yields
the identifications

$$\eqalign{
\Psi_1(x) & \rightarrow (2 \pi)^{-{1\over 4}} D(x, 1): e^{{1 \over 2}
\Phi(x)} : \cr
\Psi_2(x) & \rightarrow (2 \pi)^{-{1\over 4}} D(x,1): e^{-{1 \over 2}
\Phi(x)}: \cr
\Psi^*_1(x) & \rightarrow (2 \pi)^{-{1 \over 4}} D(x,-1): e^{+{1 \over 2}
\Phi(x)}: \cr
\Psi^*_2(x) & \rightarrow (2 \pi)^{-{1 \over 4}} D(x, -1): e^{-{1 \over 2}
\Phi(x)}: \cr} \eqno(5.15)
$$

We show how to derive the bosonization formula for
$\langle T\Bigl(\Psi^*_{\alpha_1} (x_1)... \Psi_{\delta_n}
(y_n)\Bigr)\rangle^V,$
following the method outlined in sections 3--4.
(Bilinear expressions in the fermion fields are bosonized
by the same method, in [23]).
We couple $\Psi$ minimally to $A$, obtaining the gauge-invariant action

$$
I(\Psi, \Psi^*, A) = \int d^2 x \{\Psi^*(\parz - \A) \Psi \} (x)
\eqno(5.16)
$$

A standard computation, due originally to Schwinger, gives

$$
e^{-S(A)} = e^{-{1 \over 2 \pi} (dA, \Delta^{-1} dA)}
= e^{-{1 \over 2\pi} (A^T,A^T)} \eqno(5.17)
$$

In (5.16) $A^T$ denotes the transverse component of $A$:

$$
A^T \equiv \delta \Delta^{-1} d A = A - d\Delta^{-1} \delta A,
\eqno(5.18)
$$

where the second equality follows from the Hodge decomposition (in a
space with  $H^1_{deR} (M)=0)$.
In $d$=1 the field $b^{(0)}$ is just a scalar real field denoted by
$\Phi$.
{}From eqs. (3.8)-(3.11) and (5.17) it follows, that

$$
e^{-\tilde S(d \Phi)=} \int {\cal D} [A] e^{-{1 \over 2 \pi}(A^T, A^T)}
e^{{1 \over 2\pi} \int A \wedge d \Phi} = e^{-{1 \over 8 \pi} (d \Phi, d
\Phi)} \eqno(5.19)
$$

Hence, the action of $b^{(0)} \equiv \Phi$ coincides with (5.5).
Next, we consider the fermion correlation functions. We note that, at
non--coinciding points, the contributions coming from the left
movers  $\Psi_L \equiv \Psi_1, \Psi^*_L \equiv \Psi^*_2$
and  the  right
movers $\Psi_R \equiv \Psi_2, \Psi^*_R \equiv \Psi^*_1$
factorize, so
that one can simply consider correlation functions of right
movers. Correlation functions of left movers are obtained
by complex conjugation. The explicit expression of the
correlation functions of free fermions in $d$=1 in the presence
of a gauge field $A$ is given by the equation (see [24])

$$\eqalign{
\langle & T \Bigl(\Psi^*_R (x_1)... \Psi^*_R (x_n) \Psi_R (y_1)... \Psi_R
(y_n) \Bigr)\rangle_A
= \det \Bigl({1 \over x_i - y_j}\Bigr)\cr
\exp & \Bigl\{\int d^2 z \Bigl[i \delta \Delta^{-1} A
(z) - ^*d \Delta^{-1} A(z)\Bigr] \Bigl[\sum^n_{i=1} \delta(z - x_i) - \sum^n_
{j=1} \delta (z-y_j)\Bigr] \Bigr\}. \cr} \eqno(5.20)
$$

We now combine (4.31) with (5.17) and (5.20) and, recalling (4.6),
we obtain that

$$\eqalign{
\langle T & \Bigl(\Psi^*_R (x_1)... \Psi_R (y_n) \Bigr)\rangle^V = \cr
= & \Xi (V)^{-1} \int {\cal D} \Phi \int {\cal D} [A] e^{-{1 \over 2\pi}
(A^T, A^T)} \det \Bigl({1 \over x_i - y_j}\Bigr) \cr
& e^{{i \over 2 \pi} \int(\delta \Delta^{-1} A \wedge d
\varphi_{{\underline x},{\underline y}} - A \wedge \varphi_{{\underline
x}, {\underline y}})}
e^{-{1 \over 2 \pi} (d \Delta^
{-1} A, d \varphi_{{\underline x}, {\underline y}})} \cr
& e^{{i \over 2 \pi}  \int A \wedge d \Phi} e^{- {1 \over 4 \pi^2} (^*(
d \Phi + \varphi_{{\underline x},{\underline y}}), V^* (d \Phi + \varphi_
{{\underline x}, {\underline y}}))} \cr}. \eqno(5.21)
$$

Using (2.8), (5.18) and (4.5) we find that

$$\eqalign{
\int & A \wedge \varphi_{{\underline x}, {\underline y}}
= (A - d \delta \Delta^{-1} A, ^*\varphi_{{\underline x},
{\underline y}}) = (\delta \Delta^{-1} d A, ^*\varphi_{{\underline x},
{\underline y}}) \cr
= & (\Delta^{-1} dA, ^*\delta \varphi_{{\underline x}, {\underline y}})
= 0 \cr} \eqno(5.22)
$$

and

$$
(d \Delta^{-1} A, d \varphi_{{\underline x}, {\underline y}}) =
(\delta d \Delta^{-1} A, \varphi_{{\underline x}, {\underline y}}) = (A^T,
\varphi_{{\underline x},{\underline y}}). \eqno(5.23)
$$

Plugging (5.22) and (5.23) in (5.21) and integrating out $A$, we obtain
the identity

$$\eqalign{
\langle  & T \Bigl(\Psi^*_R (x_1)... \Psi_R (y_n) \Bigr)\rangle^V = \cr
\Xi (V)^{-1} & \int {\cal D} \Phi e^{-{\pi \over 2} {1 \over
4 \pi^2} (^* d \Phi + i \varphi_{{\underline x}, {\underline y}}, ^*d \Phi +
i \varphi_{{\underline x}, {\underline y}})} \cr
& e^{-{1 \over 4 \pi^2}  (^*(d \Phi + \varphi_{{\underline x}, {\underline
y}}), V^*(d \Phi + \varphi_{{\underline x}, {\underline y}}))} \det
\Bigl({1 \over x_i - y_j}\Bigr)  \cr}. \eqno(5.24)
$$

By direct computation, using (4.6) the cross term in the first
exponential on the r.h.s. of (5.24) is given by

$$
- {\pi \over 2} {2 i \over 4 \pi^2} (\varphi_{{\underline x}, {\underline
y}}, ^*d \Phi) = - {i \over 4 \pi} (^* d \varphi_{{\underline x},
{\underline y}}, \Phi)
= {i \over 2} \Bigl(\sum^n_{i=1} \Phi (x_i) - \sum^n_{j=1} \Phi
(y_j) \Bigr) \eqno(5.25)
$$

and the term quadratic in $\varphi_{{\underline x}, {\underline y}}$ is
given, formally (see 5.13), by:

$$\eqalign{
&{- \pi \over 2} i^2 {1 \over 4 \pi^2} (\varphi_{{\underline x},
{\underline y}}, \varphi_{{\underline x}, {\underline y}}) =
{\pi \over 2} \Bigl(^*d \Delta^{-1} \Bigl(\sum^n_{i=1} \delta_{x_i}
- \sum^n_{j=1}
\delta_{y_j}\Bigr), ^*d \Delta^{-1} \Bigl(\sum^n_{i=1} \delta_{x_i}
- \sum^n_{j=1}
\delta_{y_j}\Bigr)\Bigr) \cr
& = {\pi \over 2} \Bigl[\sum^n_{i=1} \Delta^{-1} (x_i, x_i) + \sum^n_{j=1}
\Delta^{-1} (y_j, y_j)\Bigr] \cr
& + {1 \over 2} \Bigl[\sum_{1 \leq i, j \leq n} {\rm ln} |x_i - y_j| -
\sum_{1 \leq i < i' \leq n} {\rm ln} |x_i - x_{i'}| - \sum_{1 \leq j <
j' \leq n} {\rm ln} |y_j - y_{j'}|\Bigr] \cr}. \eqno(5.26)
$$

An identity due to Cauchy permits us to rewrite $\det
\Bigl({ 1 \over x_i - y_j}\Bigr)$ as follows:

$$\eqalign{
& \det \Bigl({1 \over x_i - y_j} \Bigr)= {\prod_{1 \leq i < i' \leq n}
(x_i - x_{i'})
\prod_{1 \leq j < j' \leq n} (y_{j'} - y_j) \over \prod_{1 \leq i, j \leq
n} (x_i - y_j)} \cr
& = {\prod_{1 \leq i < i' \leq n} |x_i - x_{i'}| \prod_{1 \leq j < j' \leq
n} |y_j - y_{j'}| \over \prod_{1 \leq i, j \leq n} |x_i - y_j|} \cdot \cr
& \prod_{1 \leq i < i' \leq n} e^{i\ \arg\ (x_i - x_{i'})} \prod_{1
\leq j \leq j' \leq n} e^{i\ \arg\ (y_{j'} - y_j)} \prod_{1 \leq i, j
\leq n} e^{- i\ \arg\  (x_i - y_j)} \cr}. \eqno(5.27)
$$

Collecting (5.24) - (5.27), inserting in (5.24) and recalling (5.7),
(4.10) and (4.11) we obtain that

$$\eqalign{
& \langle T (\Psi^*_R (x_1)... \Psi_R (y_n))\rangle^V
= \Xi (V)^{-1} \int {\cal D}
\Phi
e^{-{1 \over 8 \pi} (d \Phi, d \Phi)} \prod^n_{i=1} \Bigl(e^{{i \over 2}
\Phi (x_i)} e^{{\pi \over 2} \Delta^{-1} (x_i, x_i)}\Bigr) \cr
& \prod^n_{j=1} \Bigl(e^{-{i \over 2} \Phi (y_j)} e^{{\pi \over 2}
\Delta^{-1}
(y_j, y_j)} \Bigr) {\prod_{1 \leq i < i' \leq n} |x_i - x_{i'}|^{1 \over 2}
\prod_{1 \leq j < j' \leq n} |y_j - y_{j'}|^{1 \over 2} \over \prod_{1 \leq
i, j \leq n} |x_i - y_j|^{1 \over 2}} \cr
& e^{i \sum_{1 \leq i < i' \leq n} {\rm arg} (x_i - x_{i'})}
e^{i \sum_{1 \leq j < j' \leq n} {\rm arg} (y_j - y_{j'})}
e^{-i \sum_{1 \leq i, j \leq n} {\rm arg} (x_i - y_j)} \cr
& e^{-{1 \over 4 \pi^2} (^*(d \Phi + \varphi_{{\underline x}, {\underline y}}),
V^* (d \Phi + \varphi_{{\underline x}, {\underline y}}))} = \Xi(V)^{-1} \int
{\cal D} \Phi e^{-{1 \over 8 \pi} (d \Phi + \varphi_{{\underline x},
{\underline
y}}, d \Phi + \varphi_{{\underline x}, {\underline y}})} \cr
& e^{-{1 \over 4 \pi^2} (^*(d \Phi + \varphi_{{\underline x},
{\underline y}}, V
^*(d \Phi + \varphi_{{\underline x}, {\underline y}}))} \prod^n_{i=1}
: e^{{i \over 2} (\Phi (x_i) + \alpha_{{\underline x}, {\underline y}}
(x_i))}: \cr
& \prod^n_{j=1}: e^{-{i \over 2} (\Phi (y_j) + \alpha_{{\underline x},
{\underline y}}(y_j))}: \cr
& =(2 \pi)^{-{n \over 2}} \langle D({\underline x},
{\underline y}) \prod^n_{i=1}
: e^{{i \over 2} \Phi (x_i)} : \prod^n_{j=1}:e^{-{i \over 2} \Phi (y_j)}:
\rangle^V \cr} \eqno(5.28)
$$

where we used the formal definition of normal ordering, equation (5.6).

This is exactly the result of the standard abelian bosonization, equation
(5.14). Hence we proved that one can identify $b^{(0)}$ with the bosonic
field $\Phi$ of abelian bosonization.

\vskip 0.5truecm

{\bf 6. \ Condensed matter systems}

\vskip 0.5truecm

In this section we discuss some
applications of bosonization via gauge forms to fermionic
systems of condensed matter physics.
Let us start by defining the ``scaling limit"
$S^\star(A)$ of an effective action $S(A)$. We
replace the gauge potential $A$ by a rescaled potential

$$
A^{(\lambda)} = \lambda^{-1} A ({x \over \lambda}) \eqno(6.1)
$$

$S^\star (A)$ is defined as the coefficient of the leading
term in an asymptotic expansion of $S (A^{(\lambda)})$
around $\lambda = \infty$: (see [1] for a more complete
discussion).

Similarly, given a function $f(x)$, we define its scaling
limit, $f^\star(x)$, as
the coefficient of the leading term in an asymptotic expansion
of $f(\lambda x)$ around $\lambda= \infty$.

Remarkably, all many-body systems of non-relativistic
fermions that can be controlled analytically have the
property that $S^\star (A)$ is quadratic in $A$. Here we
describe some examples.

F) We start with the free non-relativistic fermion gas.
The euclidean action of the system is given by

$$
I(\Psi, \Psi^*) = \int d^{d+1} x \Bigl\{ \Psi^* \partial_0
\Psi + {1 \over 2m} ({\pmb\nabla} \Psi)^2 + \mu
\Psi^* \Psi \Bigr\} (x), \eqno(6.2)
$$

where $m$ is the mass of the fermions and $\mu$ the chemical
potential.

In $d$=1, the Fermi surface consists of only
two points, $\pm k_F$. Since the scaling limit is dominated
by excitations with momenta close to the Fermi surface, one
expands the momentum-space fermionic two-point Green function
around the points $\pm k_F$ on the Fermi surface, in order
to calculate its scaling limit.

The result of this analysis is that, in the scaling limit,
one can introdue quasi-particle fields $\Psi_L, \Psi^*_L$
and $\Psi_R, \Psi^*_R$ corresponding to the points -- $k_F$
and $k_F$, respectively. Then $\Psi_L, \Psi^*_L$ are the fields
describing left-moving excitations, while $\Psi_R, \Psi^*_R$
describe right movers. These excitation approximately obey
a relativistic dispersion relation, $\omega \approx |p|$,
where $\omega = E - E_F$ and $p = k - k_F$. The original
field variable $\Psi(x)$ is related to the relativistic
field variables $\Psi_L(x)$ and $\Psi_R (x)$ by

$$
\Psi (x) \simeq e^{- i k_F x^1} \Psi_R (x) + e^{i k_F x^1} \Psi_L(x).
\eqno(6.3)
$$

The euclidean action of $\Psi_L$ and $\Psi_R$ is given by

$$
I(\Psi_R, \Psi^*_R, \Psi_L, \Psi^*_L) = \int d^2 x \{\Psi^*_R
(\partial_0 + i v_F  \partial_1) \Psi_R + \Psi^*_L (\partial_0 - i v_F
\partial_1) \Psi_L \} (x). \eqno(6.4)
$$

The action (6.4) is the euclidean action of free massless Dirac fermions,
with the Fermi velocity $v_F$ playing the role of the velocity
of light. As a consequence the effective action $S^\star (A)$
determined by the action (6.4) is quadratic in $A$;
see Sect. 5. Using the
bosonization method of Sect.5, one obtains a quadratic action
for the potential $b^{(0)} \equiv \Phi$ of the conserved
$U(1)$-current, see eq. (5.19). These features of one--dimensional
systems have been known for many years [25].

More recently, it has been realized (see [26] for the
original observations) that similar ideas on the scaling limit
can be used in any dimension $d$. The sum over the two points
of the Fermi surface in (6.3) must be replaced by an integration
over a higher $(d-1)$ dimensional Fermi surface; see [27].

Let $\Omega$ be the $(d-1)$-dimensional unit sphere in
momentum space. Elements of $\Omega$ are denoted by ${\pmb\omega}$.
The extension of formula (6.3) to higher dimensions in given by

$$
\Psi (x) \simeq \int_\Omega d {\pmb\omega} e^{i k_F {\bf x}
{\cdot} {\pmb\omega}} \Psi_{\pmb\omega} (x) \eqno(6.5)
$$

and the euclidean action for the fields $\Psi_{\pmb\omega}$,
describing the scaling limit of the free Fermi gas can be formally
given as

$$
I (\{ \Psi_{{\pmb\omega}}, \Psi^*_{{\pmb\omega}} \} )\simeq
\int _\Omega d {\pmb\omega} \int d^{d+1} x \{ \Psi^*_
{{\pmb\omega}} (\partial_0 + i v_F {\pmb\omega} {\cdot} {\pmb\nabla})
e^{-\alpha({\pmb\omega} \wedge {\pmb\nabla})^2}
 \Psi_{{\pmb\omega}}\} (x). \eqno(6.6)
$$

where $\alpha$ is a positive constant.

Using (6.6), it has been shown in [2], that $S^\star (A)$ is
obtained as the integral over the set of directions, $\pm$
${\pmb\omega}$, in momentum space of contributions coming from
the degrees of freedom described by the fields $\Psi_{\pmb
\omega}$ and $\Psi_{- \pmb\omega}$, corresponding to antipodal points,
$\pm {\pmb\omega}$, on the Fermi surface.
Every such contribution just corresponds to the contribution of a
one-dimensional free fermion system and is quadratic in $A_{\pmb
\omega} \equiv (A_0, {\pmb\omega} {\cdot} {\bf A})$.
Since $S^\star (A)$ is the integral of the one-dimensional
actions $S^\star (A_{{\pmb\omega}})$ over all pairs of
points  $\pm {\pmb\omega}$ in $\Omega$, it, too, is
quadratic in $A$. Technical details of this calculation will appear
in [2].
These considerations yield an explicit expression for $S^\star (A)$.
Let $\Pi^{\mu \nu} (x, y)$ denote the vacuum polarization tensor
of a system of non-interacting fermions at zero temperature,
and let $\rho^{(1)} = \rho dx^0$, where $\rho$ is the density of the
system, then:

$$
S^\star (A) = {1 \over 2} (A, \Pi^\star A) +
i (\rho^{(1)}, A). \eqno(6.7)
$$

By invariance under $U(1)$-gauge transformations, translations,
rotations and parity, the expression for $\Pi^{\mu \nu}$ is
given, in momentum space, by

$$\eqalign{
\Pi^{ij} & (k) = \Pi_\bot (k) \Bigl(\delta_{ij} - {k^i k^j \over
{\bf k}^2}\Bigr) + \Pi_{||} (k) {k^i k^j \over {\bf k}^2} \cr
\Pi^{i0} & (k) = - \Pi_{||} (k) {k^i \over k_0} \cr
\Pi^{00} & (k) \equiv \Pi_0 (k) = \Pi_{||}(k) {{\bf k}^2 \over k^2_0},
\cr} \eqno(6.8)
$$

for $i, j = 1, ..., d$. The explicit form of $\Pi_0^\star$ in $d$=1 is
given by

$$
\Pi_0^\star (k)= \chi_0 {v_F^2 k^2_1 \over k^2_0 + v^2_F k^2_1} \eqno(6.9)
$$

and, in higher dimensions, to leading order in $|{v_F {\bf k} \over k_0}
|$ and $|{k_0 \over v_F {\bf k}}|$, we have  that

$$\eqalign{
\Pi_0^\star (k) & = (\chi_0 +
\lambda_0 |{k_0 \over v_F {\bf k}}|)
\Theta \Bigl(1 - |{k_0 \over v_F {\bf k}}| \Bigr) + \tilde \chi_0 |{v_F {\bf
k} \over k_0}|^2 \Theta \Bigl(|{k_0 \over v_F {\bf k}}| -1 \Bigr) \cr
& \Pi_\bot^\star (k) = \Bigl(\chi_\bot {\bf k}^2 + \lambda_\bot |{k_0 \over v_F
{\bf k}}|\Bigr)
\Theta \Bigl(1 - |{k_0 \over v_F {\bf k}}| \Bigr)+
\tilde \chi_0 \Theta \Bigl(|{k_0 \over v_F {\bf k}}| -1
\Bigr)  \cr} \eqno(6.10)
$$

where $\chi_0, \lambda_0, \tilde \chi_0, \chi_\bot, \lambda_\bot$ are constants
depending on  $d, m, v_F$ [29].

According to the argument given before, the
expression for $\Pi^\star$ in $d>1$ can be derived from the
integration over ${\pmb \omega}$ of one-dimensional vacuum
polarizations, $\Pi_{\pmb \omega}$, relative to the ``quasi-particle
fields" $\Psi_{\pmb \omega}, \Psi_{- \pmb \omega}$:

$$
(A, \Pi^\star A) = {1 \over 2} \int_\Omega d {\pmb \omega} (A_{\pmb
\omega},
\Pi_{\pmb \omega} A_{\pmb \omega}).
$$

Let us stress again that the quadratic nature of $S^\star (A)$
does not just follow from
a ``small $A$" approximation and dimensional analysis, but it
is the result of explicit cancellations arising from the structure
of the fermion two-point function in the scaling limit.

\vskip 0.5truecm
\underbar{Remark 6.1}. \ One can bosonize directly the scaling
limit action of the free Fermi gas expressed in terms of the ``quasi-
particle fields" $\{\Psi_{\pmb \omega}, \Psi^*_{\pmb \omega} \}, {\pmb
\omega} \in \Omega$, by introducing real scalar fields
$\{\Phi_{\pmb \omega}\}, {\pmb \omega} \in \Omega$, identifying
$\Phi_{\pmb \omega}$ with $\Phi_{- \pmb \omega}$.
The result is the Luther-Haldane bosonization [26, 28], in the euclidean
path-integral formalism. From section 5 one
can derive an explicit expression for the
fermionic fields $\Psi_{\pmb \omega}, \Psi^*_{\pmb \omega}$
in the bosonized theory. With obvious notation, we have

$$\eqalign{
\Psi_{\pm {\pmb \omega}} & (x) \rightarrow (2 \pi)^{- {1 \over 4}}
D_{\pmb \omega} (x, 1) : e^{\pm {i \over 2} \Phi_{\pmb \omega}(x)}:\cr
\Psi^*_{\pm {\pmb \omega}} & (x) \rightarrow (2 \pi)^{-{1\over 4}}
D_{\pmb \omega} (x, -1): e^{\pm {i \over 2} \Phi_{\pmb \omega}(x)}: \cr}
$$

{}From now on we omit the trivial term $i (\rho^{(1)}, A)$ in the
effective actions. This corresponds to redefining the density
${\cal J}^0 (\Psi, \Psi^*)$ by subtracting the background density $\rho$.
Furthermore we set $v_F =1$.

\vskip 0.3truecm

I) Insulators and incompressible quantum fluids form another class of
fermionic systems whose (bulk) effective action in the scaling limit,
$S^\star (A)$, is quadratic in $A$. Here incompressibility means
that the connected correlation functions of the current ${\cal J}^\mu
(\Psi, \Psi^*)$ have cluster properties better than those encountered
in systems whose large-scale physics is dominated by Goldstone
bosons. From incompressibility it follows [1] that $S^\star (A)$
is local, and, for systems, with translation, rotation and parity
invariance, it is given by

$$
S^\star (A) = {1 \over 2} \int d^{d+1}x \{g_E (dA)^{0i} (dA)_{0i}
(x) + g_B (dA)^{ij} (dA)_{ij} (x) \} \eqno(6.11)
$$

where $g_E, g_B$ are constants.

H) The quantum Hall fluids are parity-breaking, two-dimensional
incompressible systems. For Laughlin fluids with translation and rotation
invariance, it has been shown in [3,1] that $S^\star$ is the Chern-Simons
action

$$
S^\star (A) = {i \sigma_H \over 4 \pi} \int dA \wedge A \eqno(6.12)
$$

where $\sigma_H$ is the Hall conductivity, $\sigma_H = {1 \over 2 \ell +
1}$, $\ell = 0, 1, 2,...$.

S) London theory and computations based on perturbation theory
suggest that also $S^\star (A)$ for B.C.S. superconductors is
quadratic and it is given by

$$\eqalign{
S^\star (A) & = {1 \over 2} \int d^{d+1}_x \{ {1 \over \lambda_L^2}
({\bf A}^T)^2 (x) + \cr
\int d^{d+1} y & (A_0 - \partial_0 \Delta^{-1}_d
\partial_i A^i) (x) \Pi_s (x-y) (A_0 - \partial_0 \Delta^{-1}_d
\partial_j A^j) (y) \} \cr} \eqno(6.14)
$$

where $\lambda_L$ is a constant (the London penetration depth),
$$
A^T_i = A_i - \partial_i \Delta^{-1}_d \partial^j A_j
$$
with $i, j= 1,...,d, \Delta_d$
denotes the $d$-dimension laplacian and $\Pi_s$
is the scaling limit of the scalar
component of the vacuum polarization in the superconductor.
To leading order in $|{{\bf k} \over k_0}|$ and $|{k_0 \over {\bf k}}|,
\Pi_s$ in $d>1$ is given by

$$
\Pi_s (k) = {1 \over \lambda^2_L} |{{\bf k} \over k_0}|^2 \Theta
\Bigl(|{k_0 \over {\bf k}}| -1 \Bigr)+ \chi_s \Theta \Bigl(1 - |{k_0
\over {\bf k}}| \Bigr) \eqno(6.15)
$$

where $\chi_s$ is a constant depending on $d, m, v_F$ [30].
The scaling limit effective actions of the systems F, I, H, S given by
equations (6.7) (6.11), (6.12), (6.14) yield a bosonized action of the form

$$
\tilde S^\star (db) = {1 \over 8 \pi^2} (^*db, (\Pi^\star)^{-1 \ *} db)
\eqno(6.16)
$$

where, in the notation of (6.8), $\Pi^\star$ is given by

F) equations (6.9), (6.10)

I)
$$
\Pi^\star_0 (k) = g_E {\bf k}^2 ,
\Pi^\star_\bot (k) = g_B {\bf k}^2 + g_E k_0^2
$$

H)
$$
(\Pi^\star)^{\mu \nu} (k) = {i \over 2\pi \sigma_H} \epsilon^{\mu \nu \rho}
k_\rho
$$

S)
$$
\Pi^\star_0 (k) = \Pi_s(k),
\Pi^\star_\bot(k) = {1 \over \lambda_L^2} \eqno (6.17)
$$

If $\Pi$ has the form (6.8) then one finds

$$\eqalign{
& \tilde S^\star (db) = {1 \over 8\pi^2} \int d^{d+1} x d^{d+1} y \Bigl\{(^*
db)^i (x) [(\Pi_\bot^\star)^{- {1 \over 2}} \cr
& (\delta_{ij} - \partial_i
\Delta^{-1}_d \partial_j) (\Pi_\bot^\star)^{- {1\over 2}}] (x,y)
(^*db)^j (y) \Bigr\}. \cr} \eqno(6.18)
$$

In particular, in $d=2$ $b$ is a 1- form and (6.18) simplify to

$$\eqalign{
& \tilde S^\star(db) = {1 \over 8 \pi^2}
\int d^{2+1} k \Bigl\{(\Pi_\bot^\star)^{-1} (k){\bf k}^2 \cr
& (b_0 - k_0 {{\bf k \cdot b} \over
{\bf k}^2})^2 + (\Pi_0^\star)^{-1}(k) {\bf k}^2 ({\bf b}^T)^2 \Bigr\}.
\cr}\eqno(6.19)
$$

For Laughlin fluids, the dual action is given by

$$
\tilde S^\star (db) = - {i \over 4\pi \sigma_H} \int b \wedge db.
\eqno(6.20)
$$

\vskip 0.5truecm
\underbar{Duality in two-dimensional systems}
\vskip 0.5truecm
Note that, in two space dimensions, $A$ and $b$ are one 1-forms, and from
formula (6.7),(6.10), (6.11), (6.14), (6.16), (6.17), (6.19), (6.20)
it follows that the actions $S^\star (A)$
and $\tilde S^\star (db)$ are related by a remarkable ``duality":

$$\eqalign{
S^\star (b) |_I & \propto \tilde S^\star (db) |_S, \cr
S^\star (b) |_S & \propto \tilde S^\star (db) |_I, \cr
S^\star (b) |_H & \propto \tilde S^\star (db) |_H, \cr}\eqno(6.21)
$$

with $(g_E, g_B)$ corresponding to $(\lambda_L^2, \chi^{-1}_S)$,
and $\sigma_H$ going to $\sigma_H^{-1}$.

In particular, since $S^\star (A) |_I$ is the Maxwell action, it
follows that $\tilde S^\star (db)|_S$ describes a massless mode,
the Goldstone
boson of the superconducting state with broken gauge invariance,
known as the Anderson - Bogoliubov mode. By Kramers-Wannier duality,
this mode can also be described by an angular variable which is a
free field.

\vskip 0.5truecm
\underbar{Disorder fields for Laughin fluids}
\vskip 0.5truecm
Some care is needed in defining the disorder fields whose expectation
values are proportional to fermion Green functions of Hall fluids
in the scaling limit. In fact, the naive definition guessed from
(4.9), (4.10),

$$
``< D ({\underline x}, {\underline q}) >^\star = \Xi^{-1} \int {\cal D} [b]
e^{i{\sigma_H \over 4\pi} \int (b+ \tilde \alpha_{{\underline x},
{\underline q}}) \wedge (db + \varphi_{{\underline x}; {\underline q}})}
= \Xi^{-1} \int_{{\cal A}^1_{{\underline x}; {\underline q}}} {\cal D}
[\tilde b] e^{i{\sigma_H \over 4\pi} \int \tilde b \wedge f (\tilde b)}",
$$

does not make sense, since $\tilde \alpha_{{\underline x};{\underline q}}$
and $\tilde b$ are not well defined 1-forms on
\hfill\break $M_{{\underline x}} \equiv
{\bf R}^{2+1} \backslash \{{\underline x}\}$.
In more mathematical terms, one observes that the definition of the
Chern--Simons action for connections $\tilde b$ on a non--trivial
bundle ${\cal A}^1_{{\underline x}; {\underline q}}$ (see remark 4.1) requires
the specification of a reference connection $\tilde b_0 \in {\cal
A}^1_{{\underline x}; {\underline q}}$. With $b = (\tilde b - \tilde b_0)
\in \Lambda^1 (M_{{\underline x}})$, the Chern-Simons action is given
by [9]:

$$
S_{C.S.} (b, \tilde b_0)= \int b \wedge db + 2 b \wedge f (\tilde b_0).
\eqno(6.22)
$$

[In the following we adapt the procedure of [10], where a more detailed
discussion can be found. With obvious modifications, our construction can
be extended to more general Chern-Simons gauge theories.]

In order to define the disorder field proportional to a 2-point
fermion function, we choose $\tilde b_0$ as follows: let $E^{\pm}$
denote a 2-form in ${\bf R}^2$ with support in a cone ${\cal C}$ ,
with apex at $0$ and contained in the positive-(negative) time half-space,
satisfying

$$
dE^{\pm} = ^*\delta_0,\eqno(6.23)
$$

and denote by $E_x$ its translation by $x$.

Then, for $x_1,..., x_2$ in the positive--time half--space
and $x_{r+1},..., x_{2n}$ in the negative-time half-space,
with $x^0_i < x^0_{i+1}$, and for a set of charges, $q_1,...,q_{2n}$
with  $\sum_{i=1} q_i = 0, |q_i| =1,$ we define
$\tilde b_0 \equiv \tilde \alpha_{{\underline x}; {\underline
q};{\underline E}}$ by

$$
f(\tilde \alpha_{{\underline x}; {\underline q}; {\underline E}})\equiv
\varphi_{{\underline x};{\underline q}; {\underline E}} = \sum^r_{i=1}
q_i E_{x } + \sum^{2n}_{j=r+1} q_j E \eqno(6.24)
$$

and assume that the cones $\{{\cal C}_{{\underline x}}\}$ ``join at infinity",
(see [10]). Since $M_{{\underline x}}$ has a boundary, $\{ {\underline x}
\}$, $S_{C.S.} (b, \tilde b_0)$ is not gauge-invariant and we need a
compensating term in the action to restore gauge invariance; this term
is obtained
by adding to $\varphi_{{\underline x}; {\underline q}; {\underline E}}$
a two-form $j_{{\underline x}; {\underline q}}$ which is a sum of
dual currents with support in a set of straight lines joining
each point $x_i$ to its projection on to the time-zero plane,
$(0, {\bf x}_i)$ and carrying charge $q_i$. The expectation value of the
disorder field, $D({\underline x},{\underline q}, {\underline E}),$
is defined by

$$
< D({\underline x}, {\underline q},{\underline E}) >^\star =
$$
$$
= \cases{\Xi^{-1} \int{\cal D} [b] e^{i{\sigma_H \over 4\pi}[
S_{C.S.} (b, \tilde \alpha_{{\underline x};{\underline q};{\underline E}})
+ \int 2 b \wedge j_{{\underline x}; {\underline q}}]} & if $d
(j_{{\underline x};{\underline q}} + \varphi_{{\underline x}; {\underline
q}; {\underline E}}) = 0$ \cr
0, & otherwise \cr} \eqno(6.25)
$$

One can think the support of $j_{{\underline x};{\underline q}}$ as
representing the euclidean worldlines of static fermions created
(destroyed) at the boundary points $x_i$ where $q_i = -1 (+1)$; since
the elctric flux is conserved by gauge invariance, the elctric
flux lines spread at the end points ${\underline x}$ in shapes described
by the distributions $E_{\underline x}$.

The result of integration in eq.(6.25) is the following: let
$\alpha_{{\underline x};{\underline q};{\underline E}}$ be a 1-form
satisfing

$$
d \alpha_{{\underline x};{\underline q};{\underline E}}=
j_{{\underline x};{\underline q}} + \varphi_{{\underline x};{\underline
q};{\underline E}},\eqno(6.26)
$$

then

$$
<D({\underline x},{\underline q},{\underline E})>^\star = e^{i{\pi \over
\sigma_H} S_{C.S.} (\alpha_{{\underline x};{\underline q}; {\underline
E}})}\eqno(6.27)
$$

If the cones $\{{\cal C}_{{\underline x}}\}$ are shrunk to non-intersecting
paths $\{\gamma_{{\underline x}} \}$ and we denote the corresponding
$1$-form by $\alpha_{{\underline x}; {\underline q};{\underline \gamma}}$
instead of $\alpha_{{\underline x},{\underline q},{\underline E}}$,
then $d \alpha_{{\underline x}; {\underline q}; {\underline \gamma}}$
is dual to an oriented link and $S_{C.S.}(\alpha_{{\underline x};
{\underline q};{\underline \gamma}})$ gives the Gauss linking number of
that link.

{}From this observation one deduces that, for a Laughlin fluid
with $\sigma_H = 1/(2 \ell +1)$, $\ell = 0,1,2,...,$ the particles
described by the Green functions $<D({\underline x,} {\underline q},
{\underline E})>^\star$ are fermions, because if two arguments,
$x_i$ and $x_j$, with $q_i = q_j$ are interchanged by
a smooth deformation of the paths $\gamma_{x_i}$ and $\gamma_{x_j}$
then $<D({\underline x}, {\underline q}, {\underline E})>^\star$
changes sign.

\vskip 0.5truecm
{\bf 7 \ Adding perturbations: some applications}
\vskip 0.5truecm
Let us perturb the ``reference" systems F) - S) by a density-density and/
or current-current interaction described by a perturbation term in the
action, $I_{pert} (\Psi, \Psi^*)$ given by equation (3.19) with $V =
(V_{\mu\nu})$ positive-definite. Furthermore
let us make the following perturbative assumption.

\vskip 0.3truecm
\underbar{Assumption P}. The scaling limit of the perturbed theory
coincides with the perturbation of the scaling limit of the ``reference"
theory.

Adopting definition (3.16), the bosonized action of the
perturbed theory in the scaling limit is given by

$$
\tilde S^\star_{tot} (db) = {1 \over 8 \pi^2} (^*db, (\Pi^\star)^{-1} +
V^\star) ^*db), \eqno(7.1)
$$

provided Assumption $P$ holds. From
(6.18) one derives, for example, that the two-point function
of charge density and current in the scaling limit is given by

$$
\langle {\cal J}_\mu (\Psi, \Psi^*; x) {\cal J}_\nu (\Psi, \Psi^*; y)
\rangle^\star =
4 \pi^2 \Bigl((\Pi^\star)^{-1} + V^\star \Bigr)^{-1}_{\mu\nu} (x, y)
\eqno(7.2)
$$

Perturbing the free theory, F, (6.19) reproduces the result of the
euclidean R.P.A. approximation in the scaling limit. This
proves that, under Assumption $P$, the R.P.A. approximation gives the
(leading term in the) scaling limit of density and current correlation
functions. Using the form (6.6) of the action for a system of free
fermions, one can argue that Assumption P holds if the interaction
$V = (V_{\mu\nu})$ is positive-definite and of very long (non-integrable)
range [2].

\vskip 0.5truecm
\underbar{Plasmon gap and Anderson-Higgs mechanism}
\vskip 0.5truecm
Equation (7.2) is useful to understand the phenomenon of ``mass
generation" via Coulomb repulsion: if one perturbs a free Fermi gas or a
B.C.S. superconductor in two or more dimensions by ($d$-dimensional)
repulsive Coulomb two-body interactions, with $V$ given by $\hat V_{00}
({\bf k})= e^2 |{\bf k}|^{-2}$, and $V_{\mu\nu} = 0$ otherwise, then the
Fourier transform of the density two-point function has quasi-particle
poles at $k_0 = \pm i M, M>0$, as $|{\bf k}| \searrow 0$. [For perturbations
described by a two-body potential $V_{00} ({\bf x})$ decaying faster than
the Coulomb potential, the denominator of the two point function in Fourier
transform vanishes, as ${\bf k}, k_0 \searrow 0$].

For a perturbation of the free theory (F), $M$ is the \underbar{plasmon
gap}. The fact that $M$ is strictly positive, has been interpreted in
[31] as the result of a ``generalized Goldstone theorem" applicable
in the presence of Coulomb forces.

For superconductor,a similar phenomenon has first been analysed by
Anderson [32]. From
(7.2) and (6.10), (6.15),and for $V_{00} = {e^2 \over |{\bf k}|^2}$
one obtains

$$
\Bigl(\langle{\cal J}^0  (\Psi, \Psi^*; k) {\cal J}^0 (\Psi, \Psi^*; -k)
\rangle^V\Bigr)^\star
{\mathop{\sim}\limits_{|{\bf k} \backslash k_0| \rightarrow 0}
{4 \pi^2 \over c} |{k_0 \over {\bf k}}|^2
+ {e^2 \over |{\bf k}|^2}} =
$$
$$
{4 \pi^2 |{\bf k}|^2 \over c(k^2_0 + M^2)},
\eqno(7.3)
$$

where $M= e^2/c, c = \tilde \chi_0^{-1}$ for F, $\lambda_L^{-2}$
for S.

As remarked above for perturbations of a system of
free electrons satisfying Assumption P, the R.P.A.
approximation is exact in the scaling limit. This
explains why the R.P.A. value of the plasmon gap coincides
with the exact value obtained by a non-perturbative analysis in [31].

For the superconducting theory one can go one step
further then if one uses a form for $\Pi^\star_0 \equiv \Pi_s$ that
correctly interpolates between the behaviour at small $|{{\bf k} \over
k_0}|$ and at large $|{{\bf k} \over k_0}|$ described in (6.15), namely

$$
\Pi_s = {{\bf k^2} \over \lambda^2_L k^2_0 + \chi^{-1}_s {\bf k}^2}
\eqno(7.4)
$$

The vacuum polarization tensor $\Pi^{\mu \nu}$ defined in eq. (6.8),
with $\Pi_{\bot} = \Pi^\star_{\bot}$  and $\Pi_0 = \Pi^\star_0 =
\Pi_s$ as in (6.17), (6.15) describes a superconductor in the scaling limit
as system of non-interacting, massless $U(1)$ Goldstone bosons.
Taking Assumption P for granted, we now wish to study the effect
of two-body Coulomb repulsion on the quasi-particle spectrum of a
superconductor.

Let us first do the calculation for a two-dimensional system. By
formula (6.19) the action $\tilde S^\star_{tot}(db)$ is then given by

$$
\tilde S^\star_{tot} (db) = {1 \over 8\pi^2} (b, \tilde \Pi b), \eqno(7.5)
$$

where $\tilde \Pi^{\mu\nu}$ is given by eq. (6.8), with

$$
\tilde \Pi_{\bot} = \lambda^2_L |{\bf k}|^2 \chi^{-1}_s k^2_0 + e^2,
\tilde \Pi^{00}=
\lambda^2_L {\bf k}^2 .\eqno(7.6)
$$

{}From these formulas we learn that ${\bf b}^T$ describes a \underbar{massive}
quasi-particle, with a dispersion relation given by

$$
\omega ({\bf k}) = \sqrt{\chi_s (\lambda^2_L |{\bf k}|^2 + e^2)} \eqno(7.7)
$$

These formulas can easily be generalized to systems in $d$ dimensions,
with \hfill\break $d>2$. The result is the same: the theory
describes one massive
quasi-particle with a dispersion relation given by (7.7). This can
be seen by recalling (7.2) with $V^{\mu\nu}(k) = \delta^{\mu 0}
\delta^{\nu 0} e^2 / {\bf k}^2$

The phenomenon described above is the Anderson-Higgs mechanism.
\vskip 0.5truecm
\underbar{Existence of the charge operator}.
\vskip 0.5truecm
Another issue that can be
analyzed, using equation (7.2), is the existence of the charge
operator $\hat{\cal Q}$ (see sect.4). Consider a Wilson loop
of rank $d-1$, $W_\alpha({\cal L}_{(d-1)}^R)$, with ${\cal L}^R_{(d-1)}$=
$\partial \Sigma^R_{(d)}$. Then $W_\alpha ({\cal L}^R_{(d-1)})$ determines
an operator proportional to exp $(i \alpha \hat{\cal Q} (\Sigma^R_{(d)})$,
where  $\hat{\cal Q} (\Sigma^R_{(d)})$ measures the charge contained
in the region $\Sigma^R_{(d)}$; see eq. (4.37).

Using equation (7.1) one obtains

$$
\langle W_\alpha ({\cal L}^R_{(d-1)}) \rangle^\star = e^{-{\alpha^2 \over 2}
\langle \int_{\Sigma^R_{(d)}} db (x) \int_{\Sigma^R_{(d)}}db
(y)\rangle^\star} =
$$
$$
= e^{-{\alpha^2 \over 2} (2\pi)^2 \int_{\Sigma^R_{(d)}} d b^\mu
(x) \int_{\Sigma^R_{(d)}} db^\nu(y) ((\Pi^\star)^{-1}+ V^\star)^{-1}_{\mu\nu}
(x,y)}=
$$
$$
= e^{-2 \pi^2 \alpha^2 \int_{\Sigma^R_{(d)}} d b^0 (x)
\int_{\Sigma^R_{(d)}} db^0 (y) ((\Pi^\star)^{-1}
+ V^\star)^{-1}_{00} (x,y)}. \eqno(7.8)
$$

For $V=0$, an easy computation gives

$$
\langle W_\alpha ({\cal L}^R_{(d-1)}) \rangle^\star
{\mathop{\sim}\limits_{R\rightarrow\infty}}
\cases {e^{-c|{\cal L}^R_{(d-1)}| {\rm ln} R} & F) \cr
e^{-c|{\cal L}^R_{(d-1)}|} & I) \cr
1 & H) \cr
e^{-c|{\cal L}^R_{(d-1)}| {\rm ln} R}, & S) \cr}\eqno(7.9)
$$

for some positive constants (denoted by $c$).

Hence, according to the criterion of sect.4, the charge operator $\hat{\cal Q}$
does not exist for free systems of fermions and for superconductors.
Charge density fluctuations are so strong that the limit (4.37) does not
exist.
For insulators or quantum Hall fluids, (7.9) implies instead the existence
of the charge operator which, at zero temperature, defines a
superselection rule.

According to 't Hooft duality one expects that the two-point disorder
correlation functions of the bosonized theory, for systems F and S, have at
most power low decay in spatial directions, while, for systems I and H,
they have at least exponential decay in spatial directions.
Since the fermion fields $\Psi, \Psi^*$ are proportional to disorder
fields, one infers that, for systems I and H, the two-point fermion
Green functions exhibit at least exponential decay in spatial directions.

If we add a short-range density-density perturbation, $V_{00}({\bf k})
\simeq {\rm const} > 0$, for $|{\bf k}| \approx 0$, then under
Assumption P, the results do not change for the perturbations of
systems F, I, S; for perturbed quantum Hall fluids, we obtain a perimeter
decay for $\langle W_\alpha ({\cal L}^R_{(1)})\rangle^\star$

If we add a long-range density-density perturbation, with $V_{00} ({\bf k})
= g |{\bf k}|^{-\alpha}, \hfill\break g > 0, 0 < \alpha \leq 2$,
we obtain perimeter decay also for the perturbed systems of F and S.

\vskip 0.5truecm
\underbar{The orthogonality catastrophe}
\vskip 0.5truecm
In this last sub-section we discuss in some detail an application of our
formalism to the problem of the ``orthogonality
catastrophe" for static sources in a Landau-Fermi liquid
perturbed by a repulsive density-density interactions.
``Orthogonality catastrophe" just means that the ground state of
the system is orthogonal to the ground state in the presence
of a static source, i.e., their overlap vanishes [33].

For $d>1$, we have the following heuristic picture: the injection
of a static source triggers the production of a number, divergent
in the thermodynamic limit, of particle-hole pairs of arbitrarily-low
energy near the Fermi surface, and this, leads to the
orthogonality catastrophe. An infinite number of particle-hole pairs is
produced because there are infinitely many degrees of freedom in the
vicinity of the Fermi surface. Hence in one-dimensional systems, a
different mechanism must be responsible for the orthogonality catastrophe.
In fact, for $d$=1, the density-density interaction drives
the system away from Landau liquid behaviour, and the orthogonality
catastrophe can be related to the vanishing of the wave-function
renormalization characteristic of Luttinger liquids [34].

We now show that, under  Assumption P
our formalism leads to  clean proof of the ``orthogonality
catastrophe" in all dimensions. Our proof shows that the
term responsible for a vanishing overlap is largely insensitive
to the structure of the density--density interaction in $d>1$,
but it strongly depends on the behaviour of the interaction, as
the momentum $|{\bf k}| \searrow 0$, in $d$=1.

We start by analyzing how one can express the overlap between
the ground states in the path-integral formalism.

Let $H (J)$ be the Hamiltonian of the fermionic system in the presence
of a static source J.
Assume that the bottom of the spectrum of $H(J)$ is given by an
eigenvalue, $E(J)$, corresponding to the energy of the ground state,
$|0>_J$ in the presence of the static source.
Denote by $|0>$ the ground state of the Hamiltonian $H$ of the
fermion system. Then we have that

$$
{\lim\limits_{t \rightarrow \infty}} \langle 0| e^{-t \Bigl(H(J) -
E(J) \Bigr)} |0\rangle = |\langle 0|0\rangle_J|^2 \eqno(7.10)
$$

This suggest that

$$
{\lim\limits_{t \rightarrow \infty}} {\langle 0| e^{-t H(J)} |0\rangle
\over \langle 0|
e^{-2t H(J)} |0\rangle^{1 \over 2} } = |\langle 0| 0\rangle_J| \eqno(7.11)
$$

and, indeed, (7.2) can be proved, provided the limit on the l.h.s. exists.

A standard application of the Feynman-Kac formula proves the
equality

$$
\langle 0| e^{-t H(J)} |0\rangle  = \Xi (J_t), \eqno(7.12)
$$

where $\Xi (J_t)$ is the (grand-canonical) partition
function of the system in the presence of a current
described by a 1-form $J_t$ given by

$$
J_t (x) = \cases{J \delta ({\bf x}) dx^0, & 0 \leq x^0  \leq t \cr
0, & otherwise \cr } \eqno(7.13)
$$

Equation (7.11) then gives

$$
|\langle 0|0\rangle_J | = {\lim\limits_{t \rightarrow \infty}}
\ {\Xi (J_t) \over \Xi (J_{2t})^{1 \over 2}} \eqno(7.14)
$$

Using an interaction term of the form (3.19) we conclude from eqs.
(3.20) (with the notation of section 3) and Assumption P
that

$$
\Xi  (J_t)  = \int {\cal D} \Psi {\cal D} \Psi^* e^{-[S(\Psi,
\Psi^*) +
(({\cal J} (\Psi, \Psi^*) -  J_t), V ({\cal J}
(\Psi, \Psi^*) - J_t))] }
$$

and the scaling limit of $\Xi (J_t)$ is given by

$$
\Xi (J_t)^\star  = \int {\cal D} [b]
e^{-{1 \over 2} ({{}^*db \over 2\pi}, (\Pi^\star)^{-1}
{{}^* db \over 2\pi})}
e^{-{1 \over 2} (({{}^*db \over 2\pi} + J_t), V^\star
({{}^* db \over 2 \pi}  + J_t))} =
$$
$$
= e^{-{1 \over 2} (J_t, V^\star ((\Pi^\star)^{-1} +
V^\star)^{-1}
(\Pi^\star)^{-1} J_t)} \eqno(7.15)
$$

to a potential $V$ describing a time-independent,
rotation invariant density-density interaction.
The Fourier transform of $V$ is then given by

$$
V_{\mu \nu}^\star (k) = \delta_{\mu \nu} \delta_{\alpha 0} V_0
(|{\bf k}|) \eqno(7.16)
$$

We assume that

$$
V_0(|{\bf k}|)\sim |{\bf k}|^{-\alpha} \eqno(7.17)
$$

with $0\leq\alpha < 2$. Plugging
(7.16) and (7.13) into (7.15), we obtain that

$$
\Xi (J_t)^\star = \exp \Bigl\{-{J^2 \over 2} \int^t_0 dx^0 \int^t_0 dy^0
(\Pi_0^\star
+ V_0^{-1}\ )^{-1} (x^0 - y^0, {\bf 0}) \Bigr\} \eqno(7.18)
$$

Using the explicit form of $\Pi_0^\star$, equation (6.10), in $d>1$, we
conclude that

$$\eqalign{
& - \ln \Xi(J_t)^\star = \int^t_0 d\tau \int^\tau_0 ds \int_{|{\bf k}|
< \Lambda} {d^dk \over (2 \pi)^d} \int^{|{\bf k}|}_0 {d k_0 \over 2 \pi}
{\cos k_0 s \over \chi_0 + V_0^{-1} (|{\bf k}|)+
\lambda_0 |{k_0 \over {\bf k}}|} + \cr
& \int^t_0 d\tau \int^\tau_0 ds \int^\Lambda_0 {dk_0 \over 2\pi}
\int_{|{\bf k}| < k_0} {d^d k \over (2 \pi)^d}
{\cos k_0 s \over \tilde\chi_0 {{\bf k}^2\over k_0^2}
+ V^{-1}_0 (|{\bf k}|)}, \cr} \eqno(7.19)
$$

where $\Lambda$ is some ultraviolet cutoff.

The second term in (7.19) is easily bounded uniformly
in $t$ .
In the first term we define

$$
\gamma \equiv {k_0 \over |{\bf k}|}, \chi_0 (|{\bf k}|) \equiv
\chi_0 + V_0^{-1} (|{\bf k}|) \eqno(7.20)
$$

and we rewrite it as

$$
{1 \over 2\pi}  \int^t_0 d \tau \int^1_0 {d \gamma \over \gamma}
\int_{|{\bf k}| < \Lambda} {d^d k \over (2 \pi)^d} {\sin |{\bf k}|
\gamma \tau
\over \chi_0 (|{\bf k}|) + \lambda_0 \gamma} =
$$
$$
= {1 \over 2\pi}  \int^t_0 d \tau \int_{|{\bf k}| < \Lambda}
{d^d k \over (2 \pi)^d} {1 \over \chi_0 (|{\bf k}|)} \{\sin ({\chi_0 (|{\bf
k}|)
|{\bf k}| \tau \over \lambda_0})\cdot
$$
$$
\Bigl[{\rm ci}  (|{\bf k}| \tau ({ \chi_0 (|{\bf k}|) +1
\over \lambda_0} )) - {\rm ci} \Bigl( {|{\bf k}| \tau\chi_0
(|{\bf k}|) \over \lambda_0}\Bigr)\Bigr]
- \cos \Bigl({\chi_0 (|{\bf k}|) |{\bf k}| \tau \over \chi_0}\Bigr) \cdot
$$
$$
\Bigl[{\rm si}  (|{\bf k}| \tau
({ \chi_0 (|{\bf k}|) \over \lambda_0} +1))- {\rm si} ({ |{\bf k}|
\tau \chi_0
(|{\bf k}|) \over \lambda_0})\Bigr] + {\rm si} (|{\bf k}| \tau) +
{\pi \over 2} \}
$$
$$
{\mathop{\sim}\limits_{t\rightarrow \infty}}
{1 \over 2\pi}  \int_{|{\bf k}| < \Lambda} {d^d k \over (2 \pi)^d}
{1 \over \chi_0 (|{\bf k}|)} \Bigl\{
{\lambda_0 \over \chi_0 (|{\bf k}|) |{\bf k}|} [\ln \Bigl({\chi_0 (|{\bf k}|)
+ \lambda_0 \over
\chi_0 (|{\bf k}|) |{\bf k}| t} - C]
\Bigr) +
$$
$$
t \Bigl({\pi \over 2}  -\arctan {1 \over t|{\bf k}|}\Bigr)
\Bigr\} \eqno(7.21)
$$

where ci and si are the cosine integral and the sine integral
functions, respectively.
Hence, as $t \nearrow \infty$, we have a contribution linear in $t$,
a contribution logarithmic in $t$ and a finite correction. According
to equations (7.14), (7.19) and (7.21), the overlap in $d>1$ is given by

$$
|\langle 0|0\rangle_J| \sim {\lim\limits_{t \rightarrow + \infty}}
\exp [- {|\lambda_0| J^2 \over
4 \pi} \int_{|{\bf k}| <\Lambda} {d^d k \over (2 \pi)^d}
{1 \over |{\bf k}|} {1 \over (\chi_0 +
V_0^{-1} (|{\bf k}|))^2} \ln t] = 0, \eqno(7.22)
$$

Equation (7.22) proves the orthogonality catastrophe in $d>1$. If one
analyses where the  vanishing of the overlap (7.22) comes from, one
realizes that it is due to the term

$$
\lambda_0 |{k_0 \over {\bf k}}| \Theta \Bigl(1 - |{k_0 \over {\bf k}}|
\Bigr) \eqno(7.23)
$$

in $\Pi_0^\star$, i.e., it originates from the region in the spectrum (low
$k_0$, low $|{\bf k}|$, $k_0 < |{\bf k}|$), where particle-hole excitations
near the Fermi surface dominate the large-scale physics.
If one omits the term (7.23), one can check that the overlap
$\langle 0| 0 \rangle_J$ becomes finite.
The term in $\Xi(J_t)^\star$ responsible for the vanishing overlap in $d>1$ is
essentially independent of the specific structure of the interaction $V_0
(|{\bf k}|)$ at low $|{\bf k}|$, since, to leading order in $|{\bf k}|$,
$V_0^{-1} (|{\bf k}|)$ gives a contribution negligible with respect
to $\chi_0$, for an interaction with $\alpha > 0$

Next, we discuss the overlap in one dimension. Using the expression for $\Pi_0$
in $d$=1, (equation (6.9)), we obtain

$$\eqalign{
- & \ln \Xi (J_t)^\star = \int^t_0 d\tau \int^\tau_0 ds \int^\Lambda_{-
\Lambda} {dk_1 \over 2 \pi} \int^{\Lambda'}_{- \Lambda'} {dk_0 \over 2\pi}
{(k^2_0 + k^2_1)cos k_0s \over (k^2_0 + k^2_1) V^{-1}_0 (k_1) +
\chi_0 k^2_1} \cr
= \int^t_0 & d \tau \int^\Lambda_{- \Lambda} {d k_1 \over 2 \pi}
\int^{\Lambda'}_{-\Lambda'} {d k_0 \over 2 \pi} {\sin k_0 \tau \over k_0}
[V_0 (k_1) - {\chi_0 V^2_0 (k_1) k^2_1 \over k^2_0 + k^2_1 (1 + \chi_0 V_0
(k_1))}] \cr
& {\mathop{\sim}_{\Lambda'\nearrow \infty}} \int^\Lambda_{-\Lambda}
{d k_1 \over 2 \pi} \int^t_0 d \tau [{1 \over 2} V_0 (k_1) - {1 \over 2}
{\chi_0 V^2_0 (k_1) \over (1 + \chi_0 V_0 (k_1)} (1 - e^{-|k_1|(1+ \chi_0 V_0
(k_1))^{1 \over 2} t})] \cr
= & \int^\Lambda_{-\Lambda} {dk_1 \over 2 \pi}
{1 \over 2} {V_0 (k_1) \over (1+ \chi_0 V_0 (k_1)} t + {1 \over 2}
{\chi_0 V_0^2 (k_1) \over (1+ \chi_0 V_0 (k_1))^{3 \over 2}} {1 \over |k_1|}
\Bigl(1 - e^{-|k_1| (1 + \chi_0 V_0 (k_1))^{1 \over 2}t} \Bigr) \cr}
\eqno(7.24)
$$

{}From equation (7.24) and (7.14) it follows that

$$\eqalign{
& |\langle 0|0 \rangle_J| \sim {\lim\limits_{t \rightarrow \infty} \exp [-
\int^\Lambda_{- \Lambda} {dk_1 \over 2 \pi}
{\chi_0 V_0^2 (k_1) \over (1 + \chi_0 V_0 (k_1))^{3 \over 2}} {1 \over
|k_1|} \cr
& \{ [{1 \over 2} (1 - e^{-|k_1| (1 + \chi_0 V_0 (k_1))^{1 \over 2} t})} -{1
\over 4} (1 - e^{-|k_1| (1 + \chi_0 V_0 (k_1))^{1 \over 2} 2t})\}] \cr}
\eqno(7.25)
$$

Contrary to the result in $d>1$, the way the limit (7.25) approaches
$0$ as $t \nearrow \infty$ depends on the specific structure of
$V_0 (k_1)$. If, e.g.,

$$
V_0 (k_1) {\mathop{\sim}_{k_1 \rightarrow 0}} g |k_1|^{-\alpha},
1 > \alpha > 0, g>0, \eqno(7.26)
$$

then

$$
|\langle 0| 0\rangle_J| \sim {\lim_{t \rightarrow \infty}} e^{-c (\alpha)
g^{1 \over 2 - \alpha} \chi_0^{{\alpha - 1 \over 2 - \alpha}} t^{\alpha
\over 2 - \alpha}} = 0 \eqno(7.27)
$$

where $c (\alpha)$ is a positive constant. If

$$
V_0 (k_1) {\mathop{\sim}_{k_1 \rightarrow 0}} g >0, \eqno(7.28)
$$

then

$$
|\langle 0| 0\rangle_J| \sim {\lim_{t \rightarrow \infty}} e^{-{1 \over 4 \pi}
{\chi_0 g^2 \over (1 + \chi_0 g)^{3 \over 2}} \ln t} = 0 \eqno(7.29)
$$

The behaviour (7.29) is characteristic of the Luttinger liquid
[34], and, in general, (7.25) suggests a non-Fermi liquid
character of the system.

\vskip 0.5truecm

\underbar{Remark 7.1}. \ If, instead of using the true $\Pi_0$,
given in equation (6.9), we use a $\tilde \Pi_0$ obtained by extending
to higher dimensions the form of $\Pi_0$ in $d$=1, i.e.,

$$
\tilde \Pi_0 (k) = \chi_0 {|{\bf k}|^2 \over k^2_0 + |{\bf
k}|^2},\eqno(7.30)
$$

then one recovers the results of [35].

\vskip 0.5truecm

\underbar{Remark 7.2}. \ If, in $d>1$, we compute the partition function,
$\Xi (J^{\bf v}_t)$, of the system in the presence of a current, $J^{\bf
v}_t$, describing the motion of a source with constant (euclidean)
velocity ${\bf v}$ in a fermion system with transverse current-current
interactions, given by:

$$
V_{\mu\nu}^\star (k) =
\cases{(\delta_{ij} - {k_i  k_j \over {\bf k}^2}) V_\bot (k_0, |{\bf k}|) ,
& $i, j= 1, ...,d$ \cr
0, & {\rm otherwise},\cr} \eqno(7.31)
$$

with $V_\bot (0) \not = 0$, then, using (7.15), we obtain that

$$\eqalign{
& \Xi(J^{\bf v}_t)^\star = \exp \{- {1\over 2} \int^t_0 dx^0 \int^t_0
dy^0 \int {d^{d+1} k \over (2 \pi)^{d+1}}\cr
& \Bigl({\bf v}^2 - {({\bf v} {\cdot}  {\bf k})^2 \over |{\bf k}|^2} \Bigr)
\Bigl(\Pi_\bot^\star (k) + V^{-1}_\bot (k) \Bigr)^{-1}
e^{i({\bf k} \cdot {\bf v}
+ k_0) (x^0 - y^0)} \cr} \eqno(7.32)
$$

To leading order in $|{k_0 \over {\bf k}}|$, for $|{k_0 \over {\bf k}}|<1$,
we derive from eq. (6.10) that

$$\eqalign{
& \Pi_\bot (k) = \chi_\bot |{\bf k}|^2 + \lambda_\bot
|{k_0 \over {\bf k}}|, \cr
& \Pi_0 (k) = \chi_0 + \lambda_0 |{k_0 \over {\bf k}}| \cr}
$$

Hence, for $k_0 = 0,$, ${\bf k} \searrow 0$, $\Pi_\bot$ vanishes as $|{\bf
k}|^2$,whereas $\Pi_0$ remains finite.
As a consequence, in contrast to density-density interactions
current-current interactions in $d>1$ may significantly
change the behaviour of $\Xi(J^{\bf v}_t)^\star$, suggesting that Fermi
systems in $d>1$ with long range current-current interactions could show
non-Fermi liquid behaviour.

In a two-dimensional system with Coulomb-Amp\`ere current-current interactions,
a non-Fermi liquid (``Luttinger") behaviour has been exhibited
within the eikonal approximation [36] (see also [37]).

\vfill\eject
{\bf References}
\vskip 0.3truecm
\item{[1]} J. Fr\"ohlich, U.M. Studer, Rev. Mod. Phys.
\underbar{65}, 733 (1993).

\item{[2]} J. Fr\"ohlich, R. G\"otschmann and P.A. Marchetti, to
appear.

\item{[3]} J. Fr\"ohlich, T. Kerler, Nucl. Phys. \underbar{B354},
369 (1991)

\item{[4]} F. Wegner, J.Math. Phys. \underbar{12}, 2259 (1971).

\item{[5]} G. 't Hooft, Nucl. Phys. B. \underbar{138}, 1(1978).

\item{[6]} C.P. Burgess, F. Quevedo ``Non abelian Bosonization
as Duality", preprint HEP-TH/9403173 (1994).

\item{[7]} W.M. Tulczyiew, Rep. Math. Phys. \underbar{16}, 233
(1979).

\item{[8]} P.A. Marchetti, R. Percacci, Lett. Math. Phys.
\underbar{6}, 405 (1982) and unpublished, P.A. Marchetti,
Master Thesis at I.S.A.S. - Trieste (1982).

\item{[9]} For a review, See e.g. T. Eguchi, P.G.Gilkey and A.T. Hanson
Phys. Rep. \underbar{66}, 215 (1980).

\item{[10]} J. Fr\"ohlich, P.A. Marchetti, Commun Math. Phys.
\underbar{121}, 177 (1989).

\item{[11]} P.A. Marchetti, M.Tonin, Nuovo Cimento
\underbar{63A}, 459 (1981) and
references therein; A. de Pantz, P.A. Marchetti and R. Percacci, J. Phys. A
\underbar{16}, 861 (1983).

\item{[12]} L.P. Kadanoff, H. Ceva,  Phys. Rev. B \underbar{3},
3918 (1971), E.C. Marino, J.A.  Swieca, Nucl. Phys. B \underbar{170}
[FS1], 175 (1980); E.C. Marino, B. Schroer and  J.A. Swieca, Nucl.
Phys. B \underbar{200} [FS4], 473 (1982), J. Fr\"ohlich, P.A.
Marchetti, Commun Math. Phys. \underbar{112}, 343 (1987).

\item{[13]} J. Fr\"ohlich, P.A. Marchetti, Commun Math. Phys.
\underbar{116}, 127 (1988), P.A. Marchetti Europhys. Lett.
\underbar{4}, 663 (1987).

\item{[14]} K. Osterwalder, R. Schrader, Commun. Math. Phys.
\underbar{31}, 33 (1973); \underbar{42}, 281 (1975).

\item{[15]} J. Fr\"ohlich in ``Recent developments in gauge
theories (Carg\`ese 1979), G. 't Hooft et al (eds.) New York:
Plenum Press 1980.

\item{[16]} J.Ginibre, C. Gruber Commun Math. Phys.
\underbar{11}, 198 (1968) and references therein.

\item{[17]} J. Fr\"ohlich, P.A.Marchetti, Phys. Rev. B \underbar{46},
6535 (1992).

\item{[18]} J. Ambjorn, B. Durhuus and T. Jonsson, Nucl.Phys.
B \underbar{330}, 509 (1990).

\item{[19]} T. Jaroszewicz, P.S. Kurzepa, Ann. Phys. \underbar{210},
255 (1991).

\item{[20]} See e.g. E. Seiler  ``Gauge Theories as a Problem of
Constructive Quantum Field Theory and Statistical Mechanics"
Lecture Notes in Phys. vol. \underbar{159}, Berlin, Heidelberg
New York: Springer 1982.

\item{[21]} S. Coleman, Phys. Rev. D \underbar{11}, 2088 (1975)
S. Mandelstam Phys. Rev.  D \underbar{11}, 3026 (1975)

J. Fr\"ohlich, E. Seiler, Helv. Phys. Acta \underbar{49}, 889
(1976).

\item{[22]} J. Fr\"ohlich, P.A. Marchetti, Commun Math. Phys.
\underbar{116}, 127 (1988).

\item{[23]} C.P. Burgess, F. Quevedo ``Bosonization as Duality"
preprint HEP-TH/9401105 (1994).

\item{[24]} N.K. Nielsen, K.D. Rothe and B. Schroer Nucl. Phys.
B \underbar{160}, 330 (1979).

\item{[25]}E. Lieb D. Mattis J. Math. Phys. \underbar{6},
304(1965) A. Luther. I. Peshel Phys. Rev. \underbar{B 12}, 3908 (1975).

\item{[26]} A. Luther, Phys. Rev. \underbar{B19}, 320 (1979).

\item{[27]} G. Benfatto, G. Gallavotti, J. Stat. Phys.
\underbar{59}, 541 (1990); P.W. Anderson Phys. Rev. Lett.
\underbar{64} 1839 (1990); R. Shankar, Physica A \underbar{177}
530 (1991); J. Feldman, J. Magnen, V. Rivasseau and E.
Trubowitz, Helv. Phys. Acta, \underbar{65} 679 (1992).

\item{[28]} F.D.M. Haldane, Lectures at Varenna 1992, Helv.
Phys. Acta \underbar{65} 152 (1992).

\item{[29]} See e.g. D. Pines, P.Nozi\`ere ``The Theory of
Quantum Liquids I" Benjamin New York - Amsterdam 1966;
A.L. Fetter, J.D. Walecka ``Quantum Theory of many particle
systems". Mc Graw - Hill 1971.

\item{[30]} See e.g. J.R. Schrieffer ``Theory of Superconductivity",
Benjamin - Cummings, Menlo Park 1964.

\item{[31]} G. Morchio, F. Strocchi, Ann. Phys. \underbar{170},
310 (1986).

\item{[32]} P.W. Anderson Phys. Rev. \underbar{112} 1900 (1958);
\underbar{130} 439, 1963.

\item{[33]} P.W. Anderson, Phys. Rev. \underbar{164}
352 (1967); Lectures at Spring College ``Superconductivity" -
I.C.T.P. Trieste 1992 (Chapter VI of ``Princeton R.V.B.
book" to be published).

\item{[34]} J.M. Luttinger, J. Math. Phys. \underbar{15}
609 (1963); F.D.M. Haldane J. Phys. C \underbar{14} 2585 (1981).

\item{[35]} P.A. Bares, X.G. Wen, Phys. Rev. B \underbar{48}
8636 (1993).

\item{[36]}D.V. Khveshchenko, P.C.E. Stamp  Phys. Rev. B
\underbar{49} 5227 (1994).

\item{[37]} D.V. Khveshchenko, R. Hlubina and T.M. Rice
Phys. Rev. B \underbar{48} 10766 (1993).

\bye